\documentclass[aps,pre,floatfix,showpacs,amsmath,amssymb,twocolumn]{revtex4-1}
\usepackage{graphicx}

\begin{document}
\title{Glassy Dynamics of Brownian Particles with Velocity-Dependent 
Friction}

\author{Anoosheh Yazdi}
\author{Matthias Sperl}
\affiliation{Institut f\"ur Materialphysik im Weltraum,
Deutsches Zentrum f\"ur Luft- und Raumfahrt, 51170 K\"oln, Germany}

\date{\today}
\begin{abstract}

We consider a two-dimensional model system of Brownian particles in which 
slow particles are accelerated while fast particles are damped. The motion 
of the individual particles are described by a Langevin equation with 
Rayleigh-Helmholtz velocity dependent friction. In case of noninteracting 
particles, the time evolution equations lead to a non-Gaussian velocity 
distribution. The velocity dependent friction allows negative values of 
the friction or energy intakes by slow particles which we consider as 
active motion, and also causes breaking of the fluctuation dissipation 
relation. Defining the effective temperature proportional to the second 
moment of velocity, it is shown that for a constant effective temperature 
the higher the noise strength, the lower are the number of active 
particles in the system. Using the Mori-Zwanzig formalism and the 
mode-coupling approximation, the equation of motion for the density 
auto-correlation function are derived. The equations are solved using the 
equilibrium structure factors. The integration-through-transients approach 
is used to derive a relation between the structure factor in the 
stationary state considering the interacting forces, and the conventional 
equilibrium static structure factor.

\end{abstract}

\pacs{64.70.P-,64.70.Q-,05.40.-a}

\maketitle

\section{Introduction}

An active particle is defined as a particle which has the ability to 
absorb energy from its environment or an internal source of energy and 
dissipate the energy to undertake an out of equilibrium motion 
\cite{Hatwalne2004,Romanczuk2012}. Different collections of active 
particles e. g. biological microswimmers \cite{Cates2012,Hu2015} or 
artificial self-propelled particles \cite{Ebbens2010,Herminghaus2014}, are 
considered as active systems. It has been shown by simulation and 
experiment that active systems can reach a frozen steady state where 
single particle fluctuations are arrested \cite{Schaller2011}. The 
possibility that an active system undergoes a glass transition is 
investigated and shown theoretically \cite{Berthier2013}.

Nonequilibrium systems such as sheared colloidal suspensions 
\cite{Fuchs2002,Fuchs2009} and granular matter \cite{Kranz2010,Kranz2013} 
can undergo a glass transition or melt out of the glassy state. Active 
microrheology \cite{Gazuz2009,Gazuz2013} is applied to near glass 
transition colloidal systems to probe the nonequilibrium regimes. For 
exploring the dynamics of each of the three aforementioned systems, mode 
coupling theory \cite{Goetze2009} has been extended to the far from 
equilibrium situations. In \cite{Fuchs2002,Fuchs2009}, the 
Integration-Through-Transients (ITT) method is developed and used to 
obtain the relevant correlation functions from solving the Smoluchowski 
equation. Farage \textit{et al.} \cite{Farage2014} have used ITT to 
calculate the structure factor of an active system using the Smoluchowski 
operator. Recently an extended mode coupling scheme has been derived by 
Szamel \textit{et al.} \cite{Szamel2015pre} to describe the glassy 
dynamics of athermal self-propelled particles. Nonequilibrium motion of 
active particles near the glass transition has been studied using 
different modeling methods, e. g. considering self-propulsion of a 
constant speed in the direction of the orientations of the particles and 
body forces generated by external shear flows \cite{Farage2014}, assuming 
an internal driving force \cite{Szamel2015pre} or a colored driving and 
dissipation mechanism \cite{Berthier2013}.

In many cases the motion of biological active particles is confined to a 
plane \cite{Franks2001,Dees2008} and numbers of experiments and simulated 
systems of artificial active particles are prepared in two dimensions 
\cite{Buttinoni2013,Fily2012,Render2013}. It has been shown that charged 
particles (grains) in plasma can undertake Brownian motion 
\cite{IvlevBook2012}. Dunkel \textit{et al.} \cite{Dunkel2004} have 
studied a two-dimensional layer of charged particles in plasma which is 
trapped in an external field, numerically. They modelled the charged 
particles by a Langevin equation with a velocity-dependent friction. They 
suggest that negative (active) friction can be helpful in explaining some 
effects arising in experiment, such as the higher apparent temperature of 
the grains in comparison to the surrounding plasma. One of the simple ways 
to account for an internal propulsion mechanism is introducing a velocity 
dependent friction in the Langevin equation 
\cite{Erdmann2000,Romanczuk2012}.  The Rayleigh-Helmholtz 
\cite{Rayleigh1945} model of friction considers a nonlinear velocity 
dependent friction force $-\gamma(\mathbf{v})\mathbf{v}=\alpha 
\mathbf{v}-\beta \textbf{v}^3$. The coefficient 
$\gamma(\mathbf{v})=-\alpha+\beta \textbf{v}^2=\alpha 
(-1+\textbf{v}^2/{\textbf{v}_0}^2)$ is similar to the damping coefficient 
which was used by van der Pol \cite{vanderpol1920} to describe the 
oscillations in self sustained oscillators. A self-oscillator transfers a 
non-periodic source of energy to a periodic process, which is the 
functionality various motors have \cite{Jenkins2013}. Badoual \textit{et 
al.} \cite{Badoual2002} have used the Rayleigh-Helmholtz model to describe 
the motion of molecular motors. In many other cases the Rayleigh-Helmholtz 
force has been used to model self-propulsion as a nonequilibrium Brownian 
motion \cite{Erdmann2000,Romanczuk2012,Ganguly2013}.

In this paper, we consider a two-dimensional system of $N$ Brownian 
particles. We model the motion of each particle by the Langevin equation 
with a Rayleigh-Helmholtz friction. We choose this friction because of its 
ability of modelling the pumping of energy to the slow particles, without 
any rotational or directional dependence. We develop the time evolution 
operators and from the corresponding Fokker-Planck equation, we estimate 
the steady state distributions. The mode coupling equations for the 
density correlation functions are then derived to study the dynamical 
behavior of the system near a glass transition point 
\cite{yazdi_thesis2015}. To find out about the possible structural changes 
emerging from the nonequilibrium conditions, we use the ITT formalism.

\section{Nonlinear Langevin Equation}

To describe the motion of Brownian particles with additional energy input 
or so-called activity we use the Langevin equation with a velocity 
dependent friction \cite{Romanczuk2012}
\begin{equation}\label{Eq:Langevin_nl}
 \frac{\text{d}\mathbf{p}_i}{\text{dt}}=\mathbf{F}_i-
\gamma(\mathbf{v}_i)\mathbf{p}_i+ \boldsymbol{\xi} R_i(t).
\end{equation}
The rapidly fluctuating force $\boldsymbol{\xi} R_i(t)$, with an ensemble 
average equal to zero, represents the interaction of the Brownian particle 
with the solvent molecules. The fluctuation force is a Gaussian white 
noise \cite{Uhlenbeck1930}, which conveys that the fluctuation force 
values are normally distributed but are uncorrelated in time
\begin{equation}\label{Eq:ave_R}
\begin{split}
&\langle R_i(t)\rangle=0,\\
&\langle \boldsymbol{\xi}R_i(t) \boldsymbol{\xi}R_j(t')\rangle = 
\xi^2\delta_{ij}\ \delta(t-t').
\end{split}
\end{equation}
In some regions in the phase space, the velocity dependent friction 
$\gamma(\mathbf{v}_i)$ allows for negative friction values. When friction 
is negative, the $-\gamma(\mathbf{v}_i)\mathbf{p}_i$ force pumps 
additional mechanical energy into the particle, rather than dissipating 
the energy.

\section{Time Evolution Operators}\label{liou}

The Liouville equations for a phase variable $A(\boldsymbol{\Gamma}) = 
A(\mathbf{r}_1, \mathbf{r}_2, \dots, \mathbf{r}_N, \mathbf{p}_1, 
\mathbf{p}_2, \dots, \mathbf{p}_N)$ and for a nonequilibrium distribution 
$f$ are defined as \cite{Evans1991}
\begin{equation}\label{Eq:Liov_phase}
\frac{\text{d} A(\boldsymbol{\Gamma})}{\text{d}t} = 
i\mathcal{L}A(\boldsymbol{\Gamma}),
\end{equation}
and
\begin{equation}\label{Eq:Liov_dist}
\frac{\partial f(\boldsymbol{\Gamma},t)}{\partial t} = 
-i \mathcal{L}^{\dagger} f(\boldsymbol{\Gamma},t).
\end{equation}
In these two equations, $i\mathcal{L}$ and $i \mathcal{L}^{\dagger}$ are 
the time evolution operators for phase variables and the distribution 
function, respectively. Using Eq.~(\ref{Eq:Langevin_nl}) we can derive the 
time evolution operators
\begin{equation}\label{Eq:iL}
\begin{split}
i\mathcal{L} =  
\dot{\boldsymbol{\Gamma}}\cdotp\frac{\partial}{\partial\boldsymbol{\Gamma}}
&= \sum_i\left(\frac{\textbf{p}_i}{m}\cdotp \frac{\partial}{\partial 
\textbf{r}_i} + \boldsymbol{F}_i \cdotp \frac{\partial}{\partial 
\textbf{p}_i}\right)\\
&  +\sum_i\left(\boldsymbol{\xi} R_i(t)\cdotp\frac{\partial}{\partial
\textbf{p}_i}-\frac{\gamma(\textbf{v}_i)}{m}\textbf{p}_i\cdotp
\frac{\partial}{\partial \textbf{p}_i} \right),
\end{split}
\end{equation}
and
\begin{equation}\label{Eq:iLdag}
\begin{split}
-i\mathcal{L}^\dagger = &-\dot{\boldsymbol{\Gamma}}\cdotp 
\frac{\partial}{\partial\boldsymbol{\Gamma}} - 
(\frac{\partial}{\partial\boldsymbol{\Gamma}}\cdotp 
\dot{\boldsymbol{\Gamma}})\\
= & \sum_i\left(-\frac{\textbf{p}_i}{m}\cdotp \frac{\partial}{\partial 
\textbf{r}_i}-\boldsymbol{F}_i \cdotp 
\frac{\partial}{\partial \textbf{p}_i}\right)\\
& + \sum_i\left(-\boldsymbol{\xi} R_i(t)\cdotp\frac{\partial}{\partial
\textbf{p}_i} + \frac{\gamma(\textbf{v}_i)}{m}\textbf{p}_i\cdotp
\frac{\partial}{\partial \textbf{p}_i} \right)\\
& + \sum_i \left(\frac{1}{m} \frac{\partial \gamma(\textbf{v}_i)}{\partial 
\textbf{p}_i}\cdotp \textbf{p}_i + \frac{\gamma(\textbf{v}_i)}{m}\right).
\end{split}
\end{equation}
The term $\boldsymbol{\xi} R_i(t)\cdotp \frac{\partial}{\partial 
\mathbf{p}_i}$ appears in both time evolution operators $i\mathcal{L}$ and 
$i \mathcal{L}^{\dagger}$. Since $\boldsymbol{\xi}R_i(t)$ is a stochastic 
force, for every realization the time evolution will be different. Thus 
the variables the operators will operate on do not have a direct time 
dependence, we take an average over the noise here. We follow the 
averaging procedure in \cite{Kubo1991} (see Appendix~\ref{sec:appA}), and 
assume $m = 1$ for simplicity, therefore
\begin{equation}\label{Eq:iL2}
\begin{split}
i\mathcal{L} = & \sum_i\left(\textbf{v}_i\cdotp \frac{\partial}{\partial \textbf{r}_i} 
+ \boldsymbol{F}_i \cdotp \frac{\partial}{\partial \textbf{v}_i}\right)\\
& +\sum_i\left(-\frac{1}{2} \xi^2 \frac{\partial^2}{{\partial 
\textbf{v}_i}^2} 
- \gamma(\textbf{v}_i)\textbf{v}_i\cdotp\frac{\partial}{\partial 
\textbf{v}_i}  \right),
\end{split}
\end{equation}
and
\begin{equation}\label{Eq:iLdag2}
\begin{split}
-i\mathcal{L}^\dagger 
=& \sum_i\left(-\textbf{v}_i\cdotp \frac{\partial}{\partial 
\textbf{r}_i} - \boldsymbol{F}_i \cdotp \frac{\partial}{\partial 
\textbf{v}_i}\right)\\
& + \sum_i\left(\frac{1}{2} \xi^2\frac{\partial^2}{{\partial\text{v}_i}^2} 
+ \gamma(\textbf{v}_i) \textbf{v}_i\cdotp\frac{\partial}{\partial 
\textbf{v}_i} \right)\\
& + \sum_i \left(\frac{\partial \gamma(\textbf{v}_i)}{\partial 
\textbf{v}_i}\cdotp \textbf{v}_i+\gamma(\textbf{v}_i)\right).
\end{split}
\end{equation}

\section{Distribution Function}

\begin{figure}[htb]
\begin{center}
 \includegraphics[width=0.50\columnwidth]{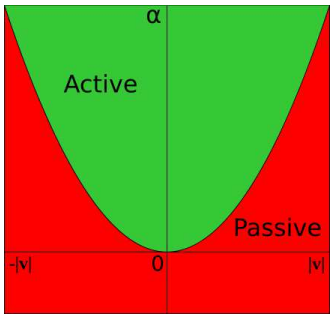}
\end{center} 
\caption{\label{fig:alpha_v} Distinct regions in the $\alpha$-$v$ plane 
which are associated with Brownian particles being active (energy intake) 
or passive (energy dissipation). The curve $\gamma(\mathbf{v}) = 
-\alpha+\mathbf{v}^2=0$ specifies the boundary of the active region.}
\end{figure}

Using the time evolution operator $-i\mathcal{L}^\dagger$ in 
Eq.~(\ref{Eq:iLdag2}), one can write the time evolution equation 
(\ref{Eq:Liov_dist}) for the distribution of one particle
\begin{equation}\label{Eq:fk}
\frac{\partial f}{\partial t} + \textbf{v}_i\cdotp \frac{\partial 
f}{\partial \textbf{r}_i} + \boldsymbol{F}_i \cdotp \frac{\partial 
f}{\partial \textbf{v}_i} = \frac{\partial }{\partial \textbf{v}_i}
\left(\gamma(\textbf{v}_i)\textbf{v}_i f + \frac{1}{2}\xi^2\frac{\partial 
f}{\partial \textbf{v}_i}\right),
\end{equation}
which is a Fokker-Planck equation. When friction is velocity dependent, 
the stationary solution of Eq.~(\ref{Eq:fk}) is only trivial when 
neglecting the interaction forces, $F_i=0$ \cite{Romanczuk2012},
\begin{equation}\label{Eq:f_s}
 f_{s}(\textbf{v}) = C \exp \left(-\frac{2}{\xi^2} \int^{\textbf{v}} 
\text{d}\textbf{v}' \gamma(\textbf{v}') \textbf{v}'\right).
\end{equation}
When $\gamma(\mathbf{v}_i) = \gamma_0=\text{Const}.$, $\xi^2 = 2 
k_\text{B} T \gamma_0$ according to the fluctuation-dissipation theorem 
\cite{Kubo1966}. In case of velocity-dependent friction, the 
fluctuation-dissipation relation does not hold which is consistent with 
the nonequilibrium situation. We consider a Rayleigh-Helmholtz model of 
friction
\begin{equation}\label{Eq:Ray_gamma_beta}
\begin{split}
  \gamma(\textbf{v}) = -\alpha+\beta \textbf{v}^2=\alpha 
(-1+\frac{\textbf{v}^2}{{\textbf{v}_0}^2}) = \beta 
(\textbf{v}^2-{\textbf{v}_0}^2),
  \end{split}
\end{equation}
where $\alpha/\beta={\textbf{v}_0}^2$ and $\beta$ takes only positive 
values. When $v < v_o$, the friction is negative and the particles receive 
energy. On the other hand, when $v > v_0$ the particles are damped due to 
the positive friction. For simplicity of analytically calculating the 
distributions we consider $\beta = 1$, so that $\alpha = {\textbf{v}_0}^2$ 
and
\begin{equation}\label{Eq:Ray_gamma}
  \gamma(\textbf{v})=-\alpha+\textbf{v}^2.
\end{equation}
We show in Fig.~\ref{fig:alpha_v} the regions in the 
$\alpha, v=|\mathbf{v}|$ plane which leads to Brownian particles being 
active (energy intake, $\gamma(\mathbf{v}) < 0$) or passive
(energy dissipation, $\gamma(\mathbf{v}) > 0$).

Considering that $\gamma(\textbf{v}) = -\alpha+\textbf{v}^2$, the 
stationary velocity distribution in Eq.~(\ref{Eq:f_s}), in terms of $D_v = 
\xi^2/2$ can be written as
\begin{equation}\label{Eq:v_distribution}
 f_\text{SR}(\textbf{v}) = C \exp \left[-\frac{1}{D_v}\left( 
\frac{\textbf{v}^4}{4}-\alpha\frac{\textbf{v}^2}{2}\right) \right].
\end{equation}
In two dimensions where $\text{d}\mathbf{v} = 2\pi~v~\text{d} v$ 
\cite{Erdmann2000},
\begin{equation}\label{Eq:1_C}
\begin{split}
 \frac{1}{C}&=2\pi\int_0^\infty \exp \left[-\frac{1}{D_v}\left( 
\frac{\textbf{v}^4}{4}-\alpha\frac{\textbf{v}^2}{2}\right) 
\right]~v~\text{d} v\\
&= \pi \sqrt{\pi D_v}~ \exp\left(\frac{\alpha ^2}{4D_v}\right) 
 \left[1+\text{erf}\left(\frac{\alpha}{2\sqrt{D_v}}\right) \right].
\end{split}
\end{equation}

\begin{figure}[htbp]
\centering
\includegraphics[width=0.9\columnwidth]{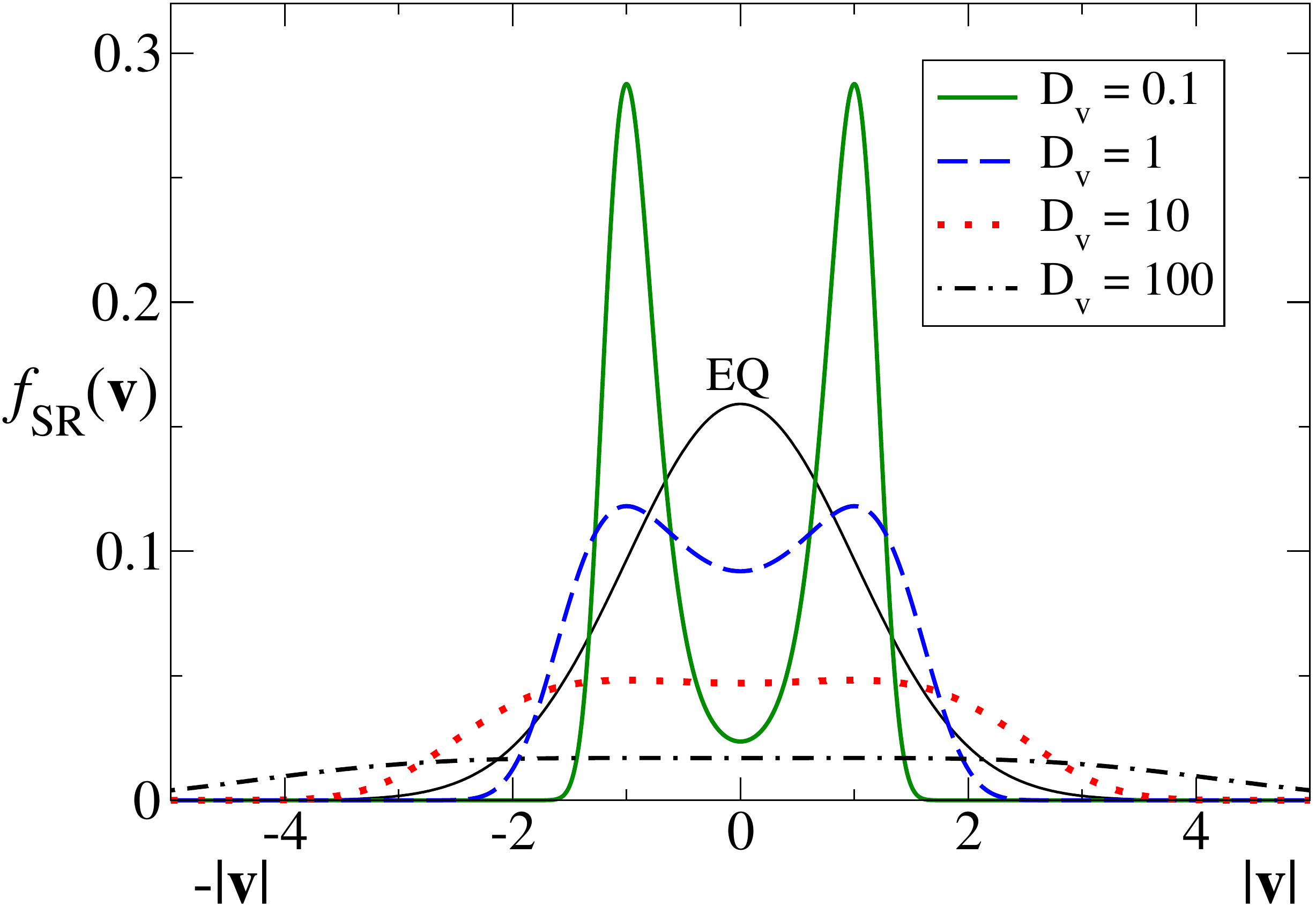}
\caption{\label{fig:fsr_1}Stationary velocity distribution for 
non-interacting Brownian particles shown in Eq.~(\ref{Eq:v_distribution}) 
for $\alpha=1$ and different values of $D_v=\xi^2/2$. The solid black line 
labeled EQ shows the normalized equilibrium Gaussian distribution 
$\exp(\alpha v^2/2 D_v)/2\pi D_v$ for $\alpha=-1$ and 
$D_v=k_\text{B}T=1$.}
\end{figure}

Figure~\ref{fig:fsr_1} shows the 2D normalized distribution $ 
f_\text{SR}(\textbf{v})$ for $\alpha=1$ and different values of $D_v$. 

\noindent%
The second, fourth and sixth moment of the velocity in two dimensions can 
be written as
\begin{equation}\label{Eq:v2}
\begin{split}
 \langle \mathbf{v}^2\rangle&= 2\pi \int_{0}^{\infty} 
f_\text{SR}(\textbf{v})~v^2~v~\text{d} v\\
  &= \alpha+ 2 \sqrt{\frac{D_v}{\pi}}
 \exp\left(-\frac{\alpha ^2}{4D_v}\right)
{\left[1+\text{erf}\left(\frac{\alpha}{2\sqrt{D_v}}\right) \right]}^{-1},
\end{split}
\end{equation}
\begin{equation}\label{Eq:v4}
\langle \mathbf{v}^4\rangle=2D_v + \alpha \langle \mathbf{v}^2\rangle,
\end{equation}
and
\begin{equation}\label{Eq:v6}
 \langle \mathbf{v}^6\rangle=2\alpha D_v+ (\alpha^2+4D_v) \langle 
\mathbf{v}^2\rangle. 
\end{equation}
These equations have been derived in Appendix~\ref{sec:appB} where we have 
also explained the slight difference between $\langle 
\mathbf{v}^2\rangle$, $\langle \mathbf{v}^4\rangle$ and what has been 
shown in \cite{Erdmann2000}. Since the velocity distribution is an even 
function, the odd moments of the velocity are zero in any dimension. The 
velocity distribution function only contains $\mathbf{v}^2 $ terms, thus 
in two dimensions: $\langle \text{v}^2_x\rangle = \langle 
\text{v}^2_y\rangle = \langle \mathbf{v}^2\rangle/2 $. We define the 
effective temperature of the system as
\begin{equation}
 k_B T_\text{eff}=\langle \text{v}^2_x\rangle = \langle 
\text{v}^2_y\rangle=\frac{\langle \mathbf{v}^2\rangle}{2}.
\end{equation}
In case of the normal Langevin equation with constant friction $\gamma_0$, 
the fluctuation-dissipation relation holds and $\xi^2/2\gamma_0 = k_BT = 
\langle\mathbf{v}^2\rangle/2$, so that there is a linear relation between 
$\langle\mathbf{v}^2\rangle$ and $\xi^2/2$. But as we can see in 
Eq.~(\ref{Eq:v2}), $\langle\mathbf{v}^2\rangle$ and $D_v = \xi^2/2$ have a 
nonlinear relation. This nonlinearity originates from the velocity 
dependent friction.

We assume that we can model the distribution of the particles with 
separating the position and velocity dependence part. For the 
Rayleigh-Helmholtz model of friction this will lead to
\begin{equation}\label{Eq:dis_total_R}
\begin{split}
 f(\{\textbf{r}_i\},\{\textbf{v}_i\}) = C & 
\exp\left(-2\frac{U(\{\mathbf{r}_i\})}{\langle \mathbf{v}^2\rangle}\right) 
\\
 &\times \exp \left[-\frac{1}{D_v} \sum_i 
\left(\frac{{\textbf{v}_i}^4}{4} - 
\alpha\frac{{\textbf{v}_i}^2}{2}\right)\right].
 \end{split}
\end{equation}
Using this distribution function in the Fokker-Planck equation and 
$D_v = \xi^2/2$ we have
\begin{equation}\label{Eq:rond_f}
\begin{split}
 \frac{\partial f}{\partial t} = \sum_i  \left(-\frac{2}{\langle 
\mathbf{v}^2\rangle} \boldsymbol{F}_i \cdot \textbf{v}_i - 
\frac{\alpha}{D_v}\boldsymbol{F}_i \cdot \textbf{v}_i
+ \frac{1}{D_v}\textbf{v}^2_i~\boldsymbol{F}_i \cdot \textbf{v}_i\right)f. 
\end{split}
\end{equation}
Multiplying the nonlinear Langevin equation (\ref{Eq:Langevin_nl}) by 
$\textbf{v}_i$ results in
\begin{equation}\label{Eq:Langevin_nl_v}
\mathbf{v}_i\cdot \frac{\text{d}\mathbf{v}_i}{\text{d}t}-\mathbf{F}_i\cdot 
\mathbf{v}_i=-\gamma(\mathbf{v}_i)\mathbf{v}^2_i+ \boldsymbol{\xi} 
R_i(t)\cdot \mathbf{v}_i,
\end{equation}
which represents the mechanical energy loss or gain of one particle in the 
system. For having the same equation in a more general form we use 
Eq.~(\ref{Eq:Liov_phase}) and (\ref{Eq:iL2}) to evaluate the time 
evolution of the variable $\sum_i \frac{\mathbf{v}^2_i}{2}$
\begin{equation}
\begin{split}
\frac{\text{d}}{\text{d} t} \sum_i \frac{\mathbf{v}^2_i}{2} &= \sum_i 
\mathbf{v}_i\cdot \frac{\text{d}\mathbf{v}_i}{\text{d}t}\\
&=i\mathcal{L}\sum_i \frac{\mathbf{v}^2_i}{2}\\
&=\sum_i\mathbf{F}_i\cdot \mathbf{v}_i - 
\sum_i\gamma(\mathbf{v}_i)\mathbf{v}^2_i+ \sum_i D_v.
\end{split}
\end{equation}
In an overdamped motion where $\text{d} \mathbf{v}_i/\text{d}t=0$ we have
\begin{equation}\label{Eq:Fv}
\begin{split}
 \sum_i\mathbf{F}_i\cdot \mathbf{v}_i &= 
\sum_i\gamma(\mathbf{v}_i)\mathbf{v}^2_i-\sum_i D_v\\
 &=-\sum_i\alpha \mathbf{v}^2_i+\sum_i \mathbf{v}^4_i-\sum_i D_v.
\end{split}
\end{equation}
We bring up that in case we did not have the nonlinear friction and 
instead we had the Langevin equation with the constant friction $\gamma_0$ 
which models the normal Brownian motion, $\sum_i\mathbf{F}_i\cdot 
\mathbf{v}_i=\sum_i\gamma_0\mathbf{v}^2_i-\sum_i \xi^2/2$ would be equal 
to zero, according to the fluctuation-dissipation relation 
$\xi^2=2k_\text{B}T \gamma_0$. But here because of the nonlinear friction 
the fluctuation-dissipation relation does not hold.

\noindent%
Replacing Eq.~(\ref{Eq:Fv}) in Eq.~(\ref{Eq:rond_f}) leads to
\begin{equation}\label{Eq:partial_f_R}
  \frac{\partial f}{\partial t}=\Lambda f,
\end{equation}
where 
\begin{equation}\label{Eq:Lambda}
\begin{split}
 \Lambda =& \left(\alpha N+\frac{2 N D_v}{\langle \mathbf{v}^2\rangle} 
\right) + \left(\frac{\alpha^2}{D_v}+\frac{2\alpha}{\langle 
\mathbf{v}^2\rangle}-1\right)\sum_i\mathbf{v}^2_i\\
&+ \left(-\frac{2\alpha}{D_v}-\frac{2}{\langle \mathbf{v}^2\rangle}\right) 
\sum_i \mathbf{v}^4_i+\frac{1}{D_v}\sum_i \mathbf{v}^6_i.
\end{split}
\end{equation}
With help of the ITT formalism, we will use $\Lambda$ in section \ref{itt} 
to write a structural relation between the stationary state at $t 
\rightarrow \infty$ and the equilibrium state.

\subsection{Probability of Finding Particles with Negative Friction 
(Active Particles)}\label{percent}

For every system having a distribution function with a specific value of 
$\alpha$ and $D_v$, which follows Eq.~(\ref{Eq:v_distribution}), the 
probability of finding particles which have a velocity less than 
$\sqrt{\alpha}$ is equal to
\begin{equation}
\begin{split}
 P_\text{active} = \int_0^{\sqrt{\alpha}} 2\pi 
f_\text{SR}(\mathbf{v})~v~\text{d} v.
 \end{split}
\end{equation}

\begin{figure}[htb]
\centering
\includegraphics[width=0.9\columnwidth]{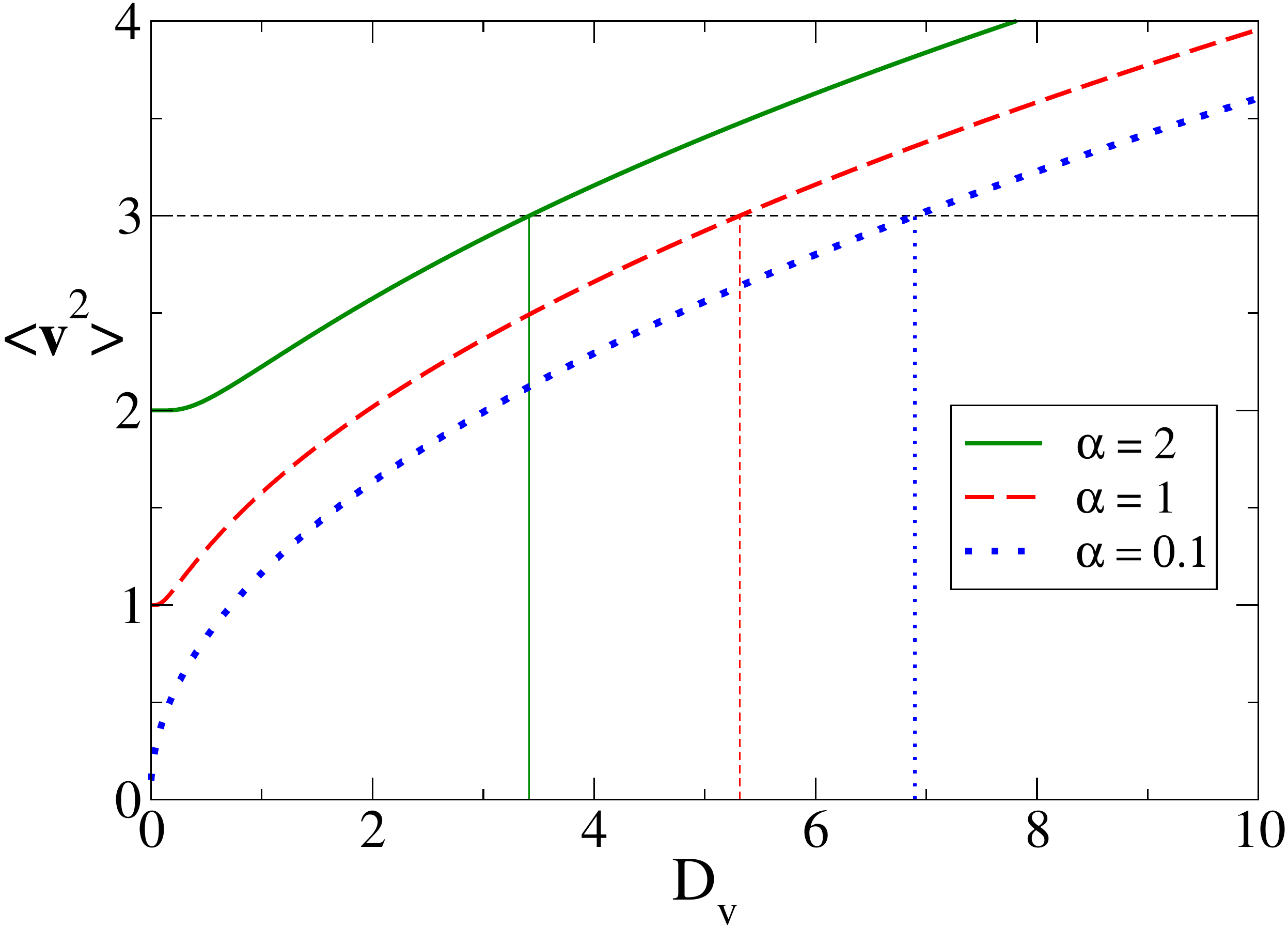}
\caption{\label{fig:v2_dv_active} Second moment of the velocity vs $D_v$ 
for three different values of $\alpha$ according to Eq.~(\ref{Eq:v2}). 
With the $\alpha$ values chosen, $\langle \mathbf{v}^2 \rangle= 2 
k_\text{B} T_\text{eff}=3$ leads to three different pairs of $(\alpha,\, 
D_v) = (0.1,6.897)$, $(1, 5.315)$ and $(2, 3.415)$.}
\end{figure}

\begin{figure}[hbt]
\centering
\includegraphics[width=0.9\columnwidth]{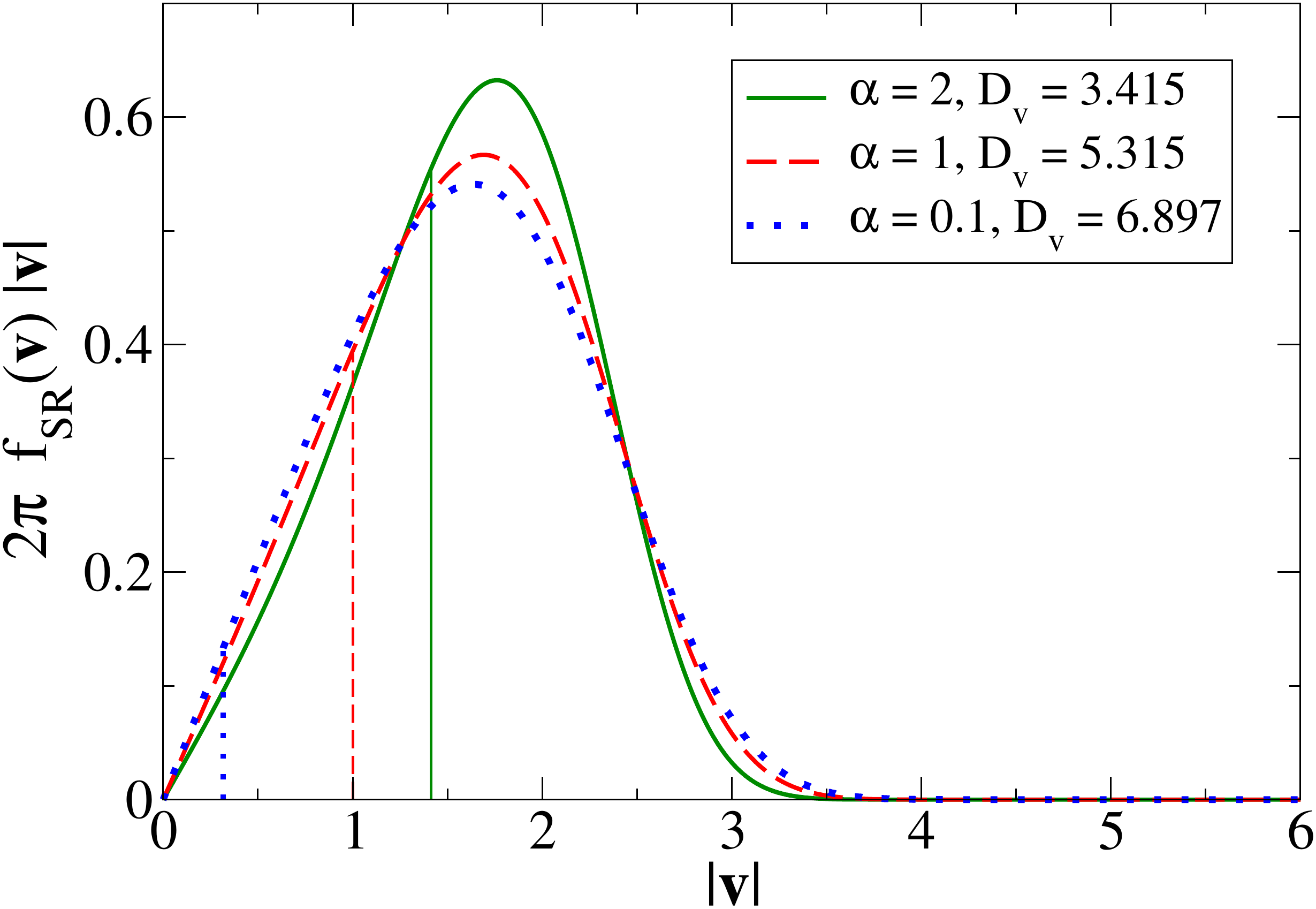}
\caption{\label{fig:p_active_Dv}Stationary velocity distributions for 
non-interacting Brownian particles shown in Eq.~(\ref{Eq:v_distribution}), 
multiplied by $2\pi v$, for different pairs of $\alpha$ and $D_v$. The 
$(\alpha, D_v)$ pairs are chosen as in Fig.~(\ref{fig:v2_dv_active}). The 
value $\sqrt{\alpha}$ is shown with vertical lines having the identical 
line style with every curve. The probability of finding particles with the 
velocity between zero and $\sqrt{\alpha}$ is equal to the area under the 
curves in that interval. This area is $0.021$, $0.288$ and $0.357$ for the 
dotted curve $(\alpha = 0.1,\, D_v=6.897)$, the dashed curve $(\alpha = 
1,\, D_v = 5.315)$ and the solid curve $(\alpha = 2,\, D_v=3.415)$, 
respectively. When temperature is constant, with increasing the $\alpha$, 
the probability of finding the particles which show activity, increases.}
\end{figure}

\noindent%
The integral can be solved as 
\begin{equation}\label{eq:n_active}
\begin{split}
 P_\text{active} &= 2\pi C \int_0^{\sqrt{\alpha}} \exp 
\left[-\frac{1}{D_v}\left(\frac{\textbf{v}^4}{4} - 
\alpha\frac{\textbf{v}^2}{2}\right) \right]~v~\text{d} v\\
&= \frac{\text{erf} \left(\frac{\alpha}{2\sqrt{D_v}}\right)}{1 
+ \text{erf}\left(\frac{\alpha}{2\sqrt{D_v}}\right)}.
 \end{split}
\end{equation}

Therefore, to compare two systems which have different values of $\alpha$ 
and $D_v$, we can use Eq.~(\ref{eq:n_active}). The larger the 
$P_\text{active}$, the larger the percentage of particles in the system 
with negative friction. As it is represented in 
Fig.~\ref{fig:v2_dv_active}, for a constant temperature $\langle 
\mathbf{v} \rangle = 2 k_\text{B} T_\text{eff} = 3$, we choose three pairs 
of $(\alpha, D_v)$. Using Eq.~(\ref{eq:n_active}), we can obtain the 
probability of finding active particles in the systems which are 
determined by these three pairs. The $P_\text{active}$ is equal to 
$0.021$, $0.288$ and $0.357$ for $(\alpha = 0.1,\, D_v = 6.897)$ , 
$(\alpha = 1,D_v = 5.315)$ and $(\alpha = 2,\, D_v = 3.415)$, 
respectively. The probability that a particle is active is 
equal to the area under the corresponding $2\pi f_\text{SR}(\mathbf{v}) v$ 
curve between zero and $v = \sqrt{\alpha}$, see 
Fig.~\ref{fig:p_active_Dv}. For a constant effective temperature, the 
larger the $\alpha$ is (or the smaller the $D_v$ is), the percentage of 
active particles in the system is higher.

\subsection{Definition of the Averages}

It will be useful for later sections to have a consistent definition of 
the ensemble averages of the product of the phase variables $A$ and 
$i\mathcal{L}B$:
\begin{equation}\label{Eq:ave_L}
 \langle A^{\ast}|i \mathcal{L}B\rangle = \int f 
A^{\ast}~i\mathcal{L}B~\text{d}\boldsymbol{\Gamma},
\end{equation}
and
\begin{equation}
 \langle -i\mathcal{L}^{\dagger}A^{\ast}|B\rangle = -\int  
\left(i\mathcal{L}^{\dagger}f A^{\ast}\right)
B~\text{d}\boldsymbol{\Gamma}.
\end{equation}
The effect of $i\mathcal{L}^{\dagger}$ on $f A^\ast$ can be evaluated as 
\cite{Evans1991},
\begin{equation}
\begin{split}
 i\mathcal{L}^{\dagger}f A^{\ast}& = \dot{\boldsymbol{\Gamma}}\cdotp 
\frac{\partial}{\partial\boldsymbol{\Gamma}}\left(f A^{\ast}
\right) + (\frac{\partial}{\partial\boldsymbol{\Gamma}}\cdotp 
\dot{\boldsymbol{\Gamma}})f A^{\ast}\\
&= f~\dot{\boldsymbol{\Gamma}}\cdotp \frac{\partial A^{\ast}}{ 
\partial\boldsymbol{\Gamma}} + A^{\ast}~\dot{\boldsymbol{\Gamma}}
\cdotp \frac{\partial f}{\partial\boldsymbol{\Gamma}} + 
A^{\ast}(\frac{\partial}{\partial\boldsymbol{\Gamma}}\cdotp 
\dot{\boldsymbol{\Gamma}})f\\
&= f~i\mathcal{L}A^{\ast}+A^{\ast}i\mathcal{L}^{\dagger}f.
\end{split}
\end{equation}
The distribution function noted in Eq.~(\ref{Eq:dis_total_R}) is not the 
stationary solution of the Fokker-Planck equation. Therefore 
$i\mathcal{L}^{\dagger}f$ is nonzero. In that case,
\begin{equation}
\begin{split}
  i\mathcal{L}^{\dagger}f A^{\ast} &= 
	f~i\mathcal{L}A^{\ast}+A^{\ast}i\mathcal{L}^{\dagger}f\\
  &= f~i\mathcal{L}A^{\ast}+A^{\ast}\Lambda f,                              
 \end{split}
\end{equation}
where $\Lambda$ is noted in Eq.~(\ref{Eq:Lambda}). Consequently, 
\begin{equation}\label{Eq:ave_Ldag2}
 \langle -i\mathcal{L}^{\dagger}A^{\ast}|B\rangle = -\int  f 
B~i\mathcal{L}A^{\ast}~
 \text{d}\boldsymbol{\Gamma}-\int A^\ast B \Lambda 
f~\text{d}\boldsymbol{\Gamma}.
\end{equation}

\section{Mori-Zwanzig Formalism}

\noindent
We consider two dynamical variables
\begin{equation}
\rho_\textbf{q}(t) = \sum_k \exp(i\textbf{q}\cdotp \textbf{r}_k(t))
\end{equation}
and
\begin{equation}
j^L_\textbf{q}(t) = \sum_k v^L_k \exp(i\textbf{q}\cdotp \textbf{r}_k(t)),
\end{equation}
where $\textbf{q} = (0,0,\text{q})$ and $L$ is the longitudinal direction 
parallel to $\textbf{q}$. The inner product of $\rho_\textbf{q}(t=0)$ with 
itself is $\langle \rho^\ast_\textbf{q}|\rho_\textbf{q}\rangle = NS_q$. 
For $j^L_\textbf{q}(t = 0)$ knowing that the odd moments of velocity are 
zero
\begin{equation}\label{Eq:j_norm}
\langle {j^L_\textbf{q}}^\ast|j^L_\textbf{q}\rangle = N\langle 
{v^L_i}^2\rangle=\frac{N}{2}\langle \mathbf{v}^2 \rangle,
\end{equation}
where $\langle \mathbf{v}^2 \rangle$ follows Eq.~(\ref{Eq:v2}). Here we 
have used the fact that the velocity distribution, 
Eq.~(\ref{Eq:v_distribution}), depends on the velocity merely through 
$|\mathbf{v}|$. So the average of the longitudinal component of the 
velocity is equal to the average of the transverse component and in two 
dimensions
\begin{equation}
\langle {v^L}^2\rangle = \langle {v^T}^2\rangle = \frac{1}{2}\langle 
\mathbf{v}^2 \rangle.
\end{equation}
In the following we use the Mori-Zwanzig formalism \cite{RZwanzig1960}, 
using the following projection operators
\begin{equation}
\begin{split}
 \mathcal{P} &= A_1\langle A_1^{\ast}|\dots \rangle +A_2\langle 
A_2^{\ast}|\dots \rangle\\
&= \frac{1}{N S_q}~\rho_\textbf{q}~\langle \rho_\textbf{q}^{\ast}|\dots 
\rangle + \frac{2}{N \langle \mathbf{v}^2 \rangle}~j^L_\textbf{q}~\langle 
{j^L_\textbf{q}}^{\ast}|\dots \rangle,
\end{split}
\end{equation}
and $\mathcal{Q} = 1 - \mathcal{P}$, where $\langle A_1^{\ast}| 
A_1\rangle$ and $\langle A_2^{\ast}| A_2\rangle = 1$. Then the equation of 
motion for the correlation function can be written as \cite{Goetze2009}
\begin{equation}\label{Eq:Mori_Zwanzig}
\left(z\textbf{I}+\boldsymbol{\Omega} - \textbf{M}\right) 
\textbf{Y}(z) = -\textbf{I},
\end{equation}
where
\begin{equation}
 Y_{nm}(z) = \langle A_n^{\ast}|\tilde{A}_m(z)\rangle,
\end{equation}
\begin{equation}
 \Omega_{nm}=\langle A_n^{\ast}|\mathcal{L} A_m\rangle,
\end{equation}
and 
\begin{equation}
M_{nm} = \langle A_n^{\ast}|\mathcal{L} \mathcal{Q} 
(z + \mathcal{Q}\mathcal{L}\mathcal{Q})^{-1} 
\mathcal{Q}\mathcal{L}A_m\rangle. \end{equation}
The $\tilde{A}_m(z) = i\int_0^{\infty} \text{d}t \exp(izt) A(t)$ is a 
Laplace transform of $A_m(t)$. With use of Eq.~(\ref{Eq:iL2}), since 
$\langle v^L \rangle = 0$, $\Omega_{11} = \frac{1}{N S_q} 
\langle\rho_\textbf{q}^{\ast}|\mathcal{L}\rho_\textbf{q} \rangle = 0$. 
From Eq.~(\ref{Eq:ave_L}) and (\ref{Eq:iL2}) we have
\begin{equation}\label{Eq:Omega_21}
 \begin{split}
  \Omega_{21}& = \frac{1}{iN \sqrt{S_q \langle \mathbf{v}^2 
\rangle/2}}\langle{j^L_\textbf{q}}^{\ast}|i\mathcal{L}\rho_\textbf{q} 
\rangle\\
  &= \frac{\sqrt{2}}{iN \sqrt{S_q \langle \mathbf{v}^2 \rangle}}
	\int \text{d}\boldsymbol{\Gamma}~f_{} \sum_k {v}_k^L 
		\exp{(-i\textbf{q}\cdotp \textbf{r}_k)}\\ 
 &\times \sum_i \textbf{v}_i\cdotp\frac{\partial}{\partial \textbf{r}_i}
 \left(\sum_{k'} \exp{(i\textbf{q}\cdotp \textbf{r}_{k'})}\right)\\
 &= \text{q}\sqrt{\frac{\langle \mathbf{v}^2 \rangle}{2S_q}}.
 \end{split}
\end{equation}
To evaluate $\Omega_{12}$ we note
\begin{equation}\label{Eq:Omega_12}
 \begin{split}
  \Omega_{12} &= \frac{1}{iN}\sqrt{\frac{2} {S_q \langle \mathbf{v}^2 
\rangle}}\langle \rho^{\ast}_\textbf{q}|i\mathcal{L} 
j^L_\textbf{q}\rangle\\
   &= \frac{1}{iN}\sqrt{\frac{2} {S_q \langle \mathbf{v}^2 
\rangle}}\left[\langle \rho^{\ast}_\textbf{q}|
 \sum_i\textbf{v}_i \cdotp\frac{\partial}{\partial 
 \textbf{r}_i}j^L_\textbf{q}\rangle\right.\\
    &\left. + \langle \rho^{\ast}_\textbf{q}| 
\sum_i\mathbf{F}_i\cdot\frac{\partial}{\partial\mathbf{v}_i} 
j^L_\textbf{q}\rangle\right.\\
   &\left. -\langle \rho^{\ast}_\textbf{q}| 
\sum_i(-\alpha+\mathbf{v}^2_i)\mathbf{v}_i\cdot 
\frac{\partial}{\partial\mathbf{v}_i} j^L_\textbf{q}\rangle\right].\\
\end{split}
\end{equation}
The third term inside the brackets contains odd moments of velocity which 
are zero and
\begin{equation}\label{Eq:Omega_1_12}
 \begin{split}
 \langle \rho^{\ast}_\textbf{q}&|\sum_i\textbf{v}_i 
\cdotp\frac{\partial}{\partial \textbf{r}_i}j^L_\textbf{q}\rangle\\
   &=iq\int \text{d}\boldsymbol{\Gamma}~f_{} \sum_{i,k} {{v}_{i}^L}^2 
\exp{[i\textbf{q}\cdotp (\textbf{r}_{i}-\textbf{r}_k)]}\\
   &=iqN\frac{\langle \mathbf{v}^2 \rangle S_q}{2}.
\end{split}
\end{equation}
Also, 
\begin{equation}\label{Eq:Omega_2_12_a}
 \begin{split}
  \langle \rho^{\ast}_\textbf{q}&| 
\sum_i\mathbf{F}_i\cdot\frac{\partial}{\partial\mathbf{v}_i} 
j^L_\textbf{q}\rangle = \\
  &=\int \text{d}\boldsymbol{\Gamma}~f_{} \sum_k\exp{(-i\textbf{q}\cdotp 
\textbf{r}_k)} \left(\sum_i\textbf{F}_i \cdotp\frac{\partial}{\partial 
\textbf{v}_i}\right)\\
   &~~\times\sum_{k'} {v}_{k'}^L \exp{(i\textbf{q}\cdotp 
\textbf{r}_{k'})}\\
   &= \int \text{d}\boldsymbol{\Gamma}~f_{} \sum_k\exp{(-i\textbf{q}\cdotp 
\textbf{r}_k)} \sum_{i} {F}_{i}^L \exp{(i\textbf{q}\cdotp 
\textbf{r}_{i})}.\\
   \end{split}
\end{equation}
We use the method applied in \cite{Gazuz2013} for a related case, to 
obtain the average in Eq.~(\ref{Eq:Omega_2_12_a}). According to 
Eq.~(\ref{Eq:dis_total_R}),
\begin{equation}\label{Eq:F_f_relation}
\frac{\partial f}{\partial \mathbf{r}_i} = -\frac{2}{\langle 
\mathbf{v}^2\rangle} \frac{\partial U}{\partial\mathbf{r}_i} f = 
\frac{2}{\langle \mathbf{v}^2\rangle} \mathbf{F}_i f,
\end{equation}
and also by means of partial integration
\begin{equation}\label{Eq:partial}
\int B~\frac{\partial f}{\partial 
\mathbf{r}_i}~\text{d}\boldsymbol{\Gamma} =
- \int f~\frac{\partial B}{\partial 
\mathbf{r}_i}~\text{d}\boldsymbol{\Gamma}.
\end{equation}
Therefore
\begin{equation}\label{Eq:Omega_2_12_b}
 \begin{split}
  &\int \text{d}\boldsymbol{\Gamma}~f_{} \sum_k\exp{(-i\textbf{q}\cdotp 
\textbf{r}_k)} \sum_{i} {F}_{i}^L \exp{(i\textbf{q}\cdotp \textbf{r}_{i})}\\
    &= -\frac{\langle \mathbf{v}^2 \rangle}{2} \sum_{i}\int 
\text{d}\boldsymbol{\Gamma} f~\frac{\partial }{\partial {r}^L_i}\left( 
   \exp{(i\textbf{q}\cdotp \textbf{r}_{i})} \sum_k\exp{(-i\textbf{q}\cdotp 
\textbf{r}_k)}\right)\\
   &= -iqN\frac{\langle \mathbf{v}^2 \rangle}{2}\left(S_q-1\right).
   \end{split}
\end{equation} 
Substituting Eq.~(\ref{Eq:Omega_2_12_b}) and (\ref{Eq:Omega_1_12}) into 
(\ref{Eq:Omega_12}) leads to
\begin{equation}
\Omega_{12} = \Omega_{21} = \text{q}\sqrt{\frac{\langle \mathbf{v}^2 
\rangle}{2S_q}}\,.
\end{equation}
This results is equivalent to the case of usual Brownian motion with 
constant friction where $\langle \mathbf{v}^2\rangle=2 k_\text{B} T$.
$\Omega_{22}$ describes sound damping, and can be evaluated as
\begin{equation}\label{Eq:omega22}
 \begin{split}
  \Omega_{22} &= \frac{2}{iN \langle \mathbf{v}^2 \rangle}\langle 
{j^L_\textbf{q}}^{\ast}|i\mathcal{L} j^L_\textbf{q}\rangle\\
  &= \frac{2}{iN \langle \mathbf{v}^2 \rangle}\int 
 \text{d}\boldsymbol{\Gamma}~f_{} \sum_k {v}_k^L \exp{(-i\textbf{q}\cdotp 
\textbf{r}_k)}\\
  &~~~\times\left(-\sum_i(-\alpha + \mathbf{v}^2_i)\mathbf{v}_i\cdot 
\frac{\partial}{\partial\mathbf{v}_i} + \sum_i\mathbf{F}_i\cdot 
\frac{\partial}{\partial\mathbf{v}_i}\right)\\
   &~~~\times\sum_{k'} {v}_{k'}^L \exp{(i\textbf{q}\cdotp 
\textbf{r}_{k'})}\\
  &= \frac{1}{i\langle \mathbf{v}^2 \rangle} 
\left(\alpha\langle\mathbf{v}^2\rangle - \langle\mathbf{v}^4\rangle\right) 
+ \frac{2}{iN \langle \mathbf{v}^2 \rangle}\int 
\text{d}\boldsymbol{\Gamma}~f_{} \sum_k v^L_k~F^L_k.
  \end{split}
\end{equation}

\noindent%
Recalling from Eq.~(\ref{Eq:v4}), $\alpha\langle\mathbf{v}^2\rangle - 
\langle\mathbf{v}^4\rangle = -2D_v = -\xi^2$. Knowing that $\sum_k 
v^L_k~F^L_k = \frac{1}{2}\sum_k \mathbf{v}_k\cdot \mathbf{F}_k$, from 
Eq.~(\ref{Eq:Fv}) we obtain
\begin{equation}\label{eq:Fv_average} 
 \int \text{d}\boldsymbol{\Gamma}~f_{} \sum_k v^L_k~F^L_k = 
\frac{N}{2}\left(-\alpha\langle\mathbf{v}^2\rangle +
 \langle\mathbf{v}^4\rangle - \frac{\xi^2}{2}\right).
\end{equation}
Therefore
\begin{equation}\label{eq:omega22_Total}
\Omega_{22} = \frac{iD_v}{\langle\mathbf{v}^2\rangle} = 
\frac{i\xi^2}{2\langle\mathbf{v}^2\rangle}.
\end{equation}

Consequently, the existence of a velocity-dependent friction term in the 
Langevin equation leads to $\langle {j^L_\textbf{q}}^{\ast}|i\mathcal{L} 
j^L_\textbf{q}\rangle = iN D_v/2$, where the $D_v$ is related to the 
second and forth moment of velocity through Eq.~(\ref{Eq:v2}). However 
$\langle {\rho_\textbf{q}}^{\ast}|i\mathcal{L} \rho_\textbf{q}\rangle$ is 
zero, similar to normal Brownian motion, since the odd moments of velocity 
are zero. The elements of the $\mathbf{\Omega}$ matrix can be written as
\begin{equation}\label{Eq:omega_matrix}
\boldsymbol{\Omega}= \begin{pmatrix}
    0 &  \text{q}\sqrt{\frac{\langle \mathbf{v}^2 \rangle}{2S_q}} \\
    \text{q}\sqrt{\frac{\langle \mathbf{v}^2 \rangle}{2S_q}} & 
	\frac{i\xi^2}{2\langle\mathbf{v}^2\rangle}\\
  \end{pmatrix}. 
\end{equation}
In case of normal Brownian motion (equilibrium case) \cite{Hess1983}, 
$\sum_i\mathbf{F}_i\cdot \mathbf{v}_i=0$ and $\Omega_{22} = i\gamma_{0}$.

\section{Mode-Coupling Approximation}

For writing the complete equation of motion, Eq.~(\ref{Eq:Mori_Zwanzig}), 
we still need to know the elements of the memory kernel $M_{nm}$. We 
recall from Eq.~(\ref{Eq:Omega_21}) that $\mathcal{L}A_1 = 
\text{q}\sqrt{\frac{\langle \mathbf{v}^2 \rangle}{2S_q}} A_2$ so 
$\mathcal{Q}\mathcal{L}A_1 = 0$ and $ M_{11} = M_{21}=0$. $M_{22}$ can be 
written as
\begin{equation}
 \begin{split}
 M_{22} &= \langle A_2^{\ast}|\mathcal{L} \mathcal{Q} 
(z + \mathcal{Q}\mathcal{L}\mathcal{Q})^{-1} 
\mathcal{Q}\mathcal{L}A_2\rangle\\
 &= \langle A_2^{\ast}|\mathcal{L} \mathcal{Q} 
\exp{(it\mathcal{QLQ})}\mathcal{Q}\mathcal{L}A_2\rangle.
 \end{split}
\end{equation}
For separating the remaining fast decaying fluctuations from the slow 
memory kernel we use the projection operator $\mathcal{P}_M = 
\sum_{\textbf{k}<\textbf{p}} \rho_{\textbf{k}} \rho_{\textbf{p}} 
\frac{\langle \rho^{\ast}_{\textbf{k}} \rho^{\ast}_{\textbf{p}}|\dots 
\rangle}{\langle \rho^{\ast}_{\textbf{k}} \rho^{\ast}_{\textbf{p}}| 
\rho_{\textbf{k}} \rho_{\textbf{p}}\rangle}$. By projecting the kernel 
onto the pair modes of density, the slowly decaying parts of the memory 
kernel remain which have the longest relaxation times \cite{Hansen1986Mc}.  
We also use the first mode-coupling approximation \cite{Goetze2009}, and 
replace $\exp{(it\mathcal{QLQ})}$ with $\mathcal{P}_M\exp{(i\mathcal{L}t)} 
\mathcal{P}_M$:
\begin{equation}\label{Eq:kernel1}
\begin{split}
M_{22} \approx& \langle A_2^{\ast}|\mathcal{L} \mathcal{Q} 
\mathcal{P}^1_M\exp{(i\mathcal{L}t)} \mathcal{P}^1_M\mathcal{Q}\mathcal{L}
A_2\rangle\\
=&\frac{2}{N\langle \mathbf{v}^2 \rangle} \sum_{\textbf{k}< \textbf{p}, 
\textbf{k}'< \textbf{p}'} \frac{1}{\langle \rho^{\ast}_{\textbf{k}} 
\rho^{\ast}_{\textbf{p}}|\rho_{\textbf{k}} \rho_{\textbf{p}}\rangle\langle 
\rho^{\ast}_{\textbf{k}'} \rho^{\ast}_{\textbf{p}'}|\rho_{\textbf{k}'} 
\rho_{\textbf{p}'}\rangle}\\
& \times \langle{j^L_\textbf{q}}^{\ast}|\mathcal{LQ} \rho_{\textbf{k}'} 
\rho_{\textbf{p}'}\rangle \langle \rho^{\ast}_{\textbf{k}'} 
\rho^{\ast}_{\textbf{p}'}| \exp{(i\mathcal{L}t)}\rho_{\textbf{k}} 
\rho_{\textbf{p}}\rangle \\
& \times \langle\rho^{\ast}_{\textbf{k}} 
\rho^{\ast}_{\textbf{p}}| \mathcal{QL} j^L_\textbf{q}\rangle
\end{split}
\end{equation}
Also according to the factorization ansatz
$\langle \rho^{\ast}_{\textbf{k}} \rho^{\ast}_{\textbf{p}}| 
\rho_{\textbf{k}} \rho_{\textbf{p}}\rangle \approx 
\langle\rho^{\ast}_{\textbf{k}}|\rho_{\textbf{k}}\rangle\langle 
\rho^{\ast}_{\textbf{p}}|\rho_{\textbf{p}}\rangle
$ and $ \langle \rho^{\ast}_{\textbf{k}'} \rho^{\ast}_{\textbf{p}'}| 
\exp{(i\mathcal{L}t)}\rho_{\textbf{k}} \rho_{\textbf{p}}\rangle \approx
\delta_{\textbf{k},\textbf{k}'}\delta_{\textbf{p},\textbf{p}'}N^2S_k 
S_p\phi_{\textbf{k}}(t)\phi_{\textbf{p}}(t),
$ where $\phi_{\textbf{k}}(t) = \langle\rho^{\ast}_{\textbf{k}}| 
\exp{(i\mathcal{L}t)} \rho_{\textbf{k}}\rangle/NS_k$. We need to calculate 
two terms, the first one:
\begin{equation}\label{Eq:j_QL_rho_rho}
\begin{split}
 \langle {j^L_\textbf{q}}^{\ast}&|\mathcal{LQ} \rho_{\textbf{k}} 
\rho_\textbf{p}\rangle\\
 =& \langle {j^L_\textbf{q}}^{\ast}|(\mathcal{L} \rho_{\textbf{k}}) 
\rho_{\textbf{p}}\rangle + \langle {j^L_\textbf{q}}^{\ast}| 
\rho_{\textbf{k}} (\mathcal{L} \rho_{\textbf{p}})\rangle
 - \frac{q\langle \mathbf{v}^2 
\rangle}{2S_q}\langle\rho^{\ast}_\textbf{q}| 
\rho_{\textbf{k}}\rho_\textbf{p}\rangle\\ 
 =& \frac{\langle \mathbf{v}^2 \rangle}{2} k \langle 
\rho^\ast_{\textbf{q}-\textbf{k}}|\rho_\textbf{p}\rangle + \frac{\langle 
\mathbf{v}^2 \rangle}{2} p \langle\rho^\ast_{\textbf{q} - 
\textbf{p}}|\rho_\textbf{k}\rangle -\frac{q\langle \mathbf{v}^2 
\rangle}{2S_q}\langle\rho^{\ast}_\textbf{q}| \rho_{\textbf{k}} 
\rho_\textbf{p}\rangle\\
 =& N\frac{\langle \mathbf{v}^2 \rangle}{2} \delta_{\textbf{q},\textbf{k} 
+ \textbf{p}}(kS_p+pS_k-qS_k S_p),
 \end{split}
\end{equation}
where we used the convolution approximation 
$\langle\rho^{\ast}_\textbf{q}| \rho_{\textbf{k}} \rho_\textbf{p}\rangle 
\approx N\delta_{\textbf{q},\textbf{k} + \textbf{p}}S_q S_k S_p$. Above 
and in the following equations, $k$ and $p$ are the longitudinal 
components of $\mathbf{k}$ and $\mathbf{p}$ respectively. The second term 
to calculate is
\begin{equation}\label{Eq:rho_rho_QL_j_a}
\begin{split}
 \langle\rho^{\ast}_{\textbf{k}} \rho^{\ast}_{\textbf{p}}| \mathcal{QL} 
 j^L_\textbf{q}\rangle=&\frac{1}{i}\langle\rho^{\ast}_{\textbf{k}} 
 \rho^{\ast}_{\textbf{p}}| i\mathcal{L} j^L_\textbf{q}\rangle
 - \langle \rho^\ast_\textbf{k} \rho^\ast_\textbf{p}| 
 \rho_\textbf{q}\rangle 
 \frac{1}{NS_q}\langle\rho^\ast_\textbf{q}|\mathcal{L} 
 j^L_\textbf{q}\rangle.\\
 \end{split}
\end{equation} 
In equilibrium, $\langle\rho^{\ast}_{\textbf{k}} \rho^{\ast}_{\textbf{p}}| 
i\mathcal{L} j^L_\textbf{q}\rangle=\langle (-i\mathcal{L}^\dagger 
\rho^\ast_\textbf{k}) \rho^\ast_\textbf{p}|j^L_\textbf{q}\rangle + \langle  
\rho^\ast_\textbf{k} (-i\mathcal{L}^\dagger\rho^\ast_\textbf{p})| 
j^L_\textbf{q}\rangle$. However here we need to let the operator 
$\mathcal{L}$ act on the variable $j^L_\textbf{q}$,
\begin{equation}\label{Eq:rho_rho_L_j}
\begin{split}
 \frac{1}{i}\langle\rho^{\ast}_{\textbf{k}} \rho^{\ast}_{\textbf{p}}| 
 i\mathcal{L} j^L_\textbf{q}\rangle 
 &= \frac{1}{i}\langle\rho^{\ast}_{\textbf{k}} \rho^{\ast}_{\textbf{p}}|
 \sum_i\textbf{v}_i \cdotp\frac{\partial}{\partial \textbf{r}_i} 
 j^L_\textbf{q}\rangle\\
 &+ \frac{1}{i}\langle\rho^{\ast}_{\textbf{k}} \rho^{\ast}_{\textbf{p}}|
 \sum_i\mathbf{F}_i\cdot\frac{\partial}{\partial\mathbf{v}_i} 
 j^L_\textbf{q}\rangle\\
 &- \frac{1}{i}\langle\rho^{\ast}_{\textbf{k}} \rho^{\ast}_{\textbf{p}}|
 \sum_i(-\alpha + \mathbf{v}^2_i) \mathbf{v}_i\cdot 
\frac{\partial}{\partial\mathbf{v}_i}j^L_\textbf{q}\rangle.\\
 \end{split}
\end{equation} 
The third term is zero since the odd moments of velocity are zero. The 
first term can be written as 
\begin{equation}\label{Eq:rho_rho_L_j_a}
\begin{split}
 \frac{1}{i}\langle\rho^{\ast}_{\textbf{k}} & \rho^{\ast}_{\textbf{p}}|
 \sum_i\textbf{v}_i \cdotp\frac{\partial}{\partial \textbf{r}_i} 
 j^L_\textbf{q}\rangle\\
 &= \frac{1}{i}\langle\rho^{\ast}_{\textbf{k}} \rho^{\ast}_{\textbf{p}}|
 \sum_i\left(\textbf{v}_i \cdotp\frac{\partial}{\partial 
 \textbf{r}_i}\right) \sum_m v^L_m \exp{(i\mathbf{q}\cdot\mathbf{r}_m)} 
\rangle\\
 &= \frac{q\langle \mathbf{v}^2\rangle}{2}\langle\rho^{\ast}_{\textbf{k}} 
\rho^{\ast}_{\textbf{p}}|\rho_{\textbf{q}}\rangle.
 \end{split}
\end{equation}
With the help of Eq.~(\ref{Eq:F_f_relation}) and (\ref{Eq:partial}) the 
second term of Eq.~(\ref{Eq:rho_rho_L_j}) can be evaluated as
\begin{equation}\label{Eq:rho_rho_L_j_b}
\begin{split}
 \frac{1}{i}\langle&\rho^{\ast}_{\textbf{k}} \rho^{\ast}_{\textbf{p}}|
 \sum_i \mathbf{F}_i\cdot 
\frac{\partial}{\partial\mathbf{v}_i}j^L_\textbf{q} \rangle\\
 &= \frac{1}{i}\langle\rho^{\ast}_{\textbf{k}} \rho^{\ast}_{\textbf{p}}|
 \sum_i F^L_i \exp{(i\mathbf{q}\cdot\mathbf{r}_i)}\rangle\\
 &= -\frac{\langle \mathbf{v}^2\rangle}{2i} \sum_i \int 
 \text{d}\boldsymbol{\Gamma} f \frac{\partial }{\partial 
 {r}^L_i}\left[\exp{(i\mathbf{q}\cdot\mathbf{r}_i)} 
 \rho^{\ast}_{\textbf{k}} \rho^{\ast}_{\textbf{p}}\right]\\
 &= -\frac{\langle \mathbf{v}^2\rangle}{2i} \sum_i \int 
 \text{d}\boldsymbol{\Gamma} f \{ iq \exp{(i\mathbf{q}\cdot\mathbf{r}_i)} 
	\rho^{\ast}_{\textbf{k}} \rho^{\ast}_{\textbf{p}}\\
  &-ik \exp{[i(\mathbf{q} - \mathbf{k})\cdot\mathbf{r}_i]} 
\rho^{\ast}_{\textbf{p}} - ip \exp{[i(\mathbf{q} - 
\mathbf{p})\cdot\mathbf{r}_i]} \rho^{\ast}_{\textbf{k}} \}\\
 &= -\frac{\langle \mathbf{v}^2\rangle}{2}  
\left(q\langle\rho^{\ast}_{\textbf{k}} \rho^{\ast}_{\textbf{p}}| 
 \rho_{\textbf{q}}\rangle -\delta_{\textbf{q},\textbf{k}+\textbf{p}}NkS_p 
- \delta_{\textbf{q},\textbf{k}+\textbf{p}}NpS_k\right).
 \end{split} 
\end{equation}
By adding up the Eq.~(\ref{Eq:rho_rho_L_j_b}) to 
Eq.~(\ref{Eq:rho_rho_L_j_a}) we have
\begin{equation}
 \frac{1}{i}\langle\rho^{\ast}_{\textbf{k}} \rho^{\ast}_{\textbf{p}}| 
 i\mathcal{L} j^L_\textbf{q}\rangle =N \frac{\langle 
 \mathbf{v}^2\rangle}{2}\delta_{\textbf{q},\textbf{k}+\textbf{p}}
 (kS_p+pS_k),
\end{equation}
so
\begin{equation}\label{Eq:rho_rho_QL_j_b}
 \langle\rho^{\ast}_{\textbf{k}} \rho^{\ast}_{\textbf{p}}| \mathcal{QL} 
 j^L_\textbf{q}\rangle= N\frac{\langle \mathbf{v}^2 
 \rangle}{2}\delta_{\textbf{q},\textbf{k}+\textbf{p}}(kS_p+pS_k-qS_k S_p).
\end{equation}
Placing Eq.~(\ref{Eq:j_QL_rho_rho}) and (\ref{Eq:rho_rho_QL_j_b}) in 
Eq.~(\ref{Eq:kernel1}) leads to
\begin{equation}
\begin{split}
 M_{22} = \frac{\langle \mathbf{v}^2\rangle}{2N} \sum_{\textbf{k} 
< \textbf{p}}~\delta_{\textbf{q},\textbf{k} + \textbf{p}}\left(\frac{k 
S_p+p S_k-\text{q} S_kS_p}{S_k S_p}\right)^2\\
\times S_k S_p\phi_{\textbf{k}}(t)\phi_{\textbf{p}}(t).
 \end{split}
\end{equation}
Therefore the expression for the kernel is the same as the MCT kernel for 
conventional liquids \cite{Goetze2009} considering $\langle \mathbf{v}^2 
\rangle/2=k_\text{B} T_\text{eff}$. The effective temperature will drop 
out by defining
\begin{equation}
 m^{mc}_q= \frac{1}{\Omega_{12}^2} M^1_{22}
\end{equation}
and $m^{mc}_q$ can be written in integral form; in two dimensions 
\cite{Bayer2007},
\begin{equation}\label{Eq:int_kernel}
\begin{split}
m^\text{mct}_q = & \int \frac{d^2 k}{(2\pi)^2} \frac{\rho S_q S_p S_k}{2 
q^4} \left(\mathbf{q}\cdot\mathbf{k} c_k + \mathbf{p}\cdot\mathbf{q} 
c_p\right) ^2 \phi_k (t) \phi_p(t),
 \end{split}
\end{equation}
where $\mathbf{p} = \mathbf{q} - \mathbf{k}$, $\rho c_k=1-1/S_k$ and 
$\rho$ is the average density for $N$ particles in an area $L^2$.

\section{Equation of Motion for the Density Auto-Correlation Function}

The equation of motion following Eq.~(\ref{Eq:Mori_Zwanzig}), 
(\ref{Eq:omega_matrix}) and (\ref{Eq:int_kernel}) can be written as
\begin{equation}\label{Eq:eq_m_newton}
\begin{split}
\partial^2_t \phi_\mathbf{q}(t)&+\frac{D_v}{\langle \mathbf{v}^2 \rangle} 
\partial_t\phi_\mathbf{q}(t)+\Omega^2_q \phi_\mathbf{q}(t)\\
&+ \Omega^2_q \int_0^t \partial_{t'}\phi_\mathbf{q}(t) 
m^\text{mct}_q(t-t')dt' = 0,
\end{split}
\end{equation}
where $\phi_\mathbf{q}(t)=\phi_{11}(t)$ and $\Omega^2_q = \Omega^2_{12} = 
q^2\langle \mathbf{v}^2 \rangle/(2S_q)$. For the overdamped case, the 
equation of motion can be written as
\begin{equation}\label{Eq:eq_m_newton0}
\begin{split}
\frac{D_v}{\langle \mathbf{v}^2 \rangle \Omega^2_q} 
\partial_t\phi_\mathbf{q}(t)+\phi_\mathbf{q}(t) + \int_0^t 
\partial_{t'}\phi_\mathbf{q}(t) m^\text{mct}_q(t-t')dt' = 0.
\end{split}
\end{equation}
The equation of motion presented as Eq.~(\ref{Eq:eq_m_newton}) contains 
one more approximation in comparison to the overdamped case in 
Eq.~(\ref{Eq:eq_m_newton0}). Seeing that we have used the property of an 
overdamped motion conveyed in Eq.~(\ref{Eq:Fv}), to calculate 
$\Omega_{22}$.

\begin{figure}[htbp]
\centering
\includegraphics[width=0.9\columnwidth]{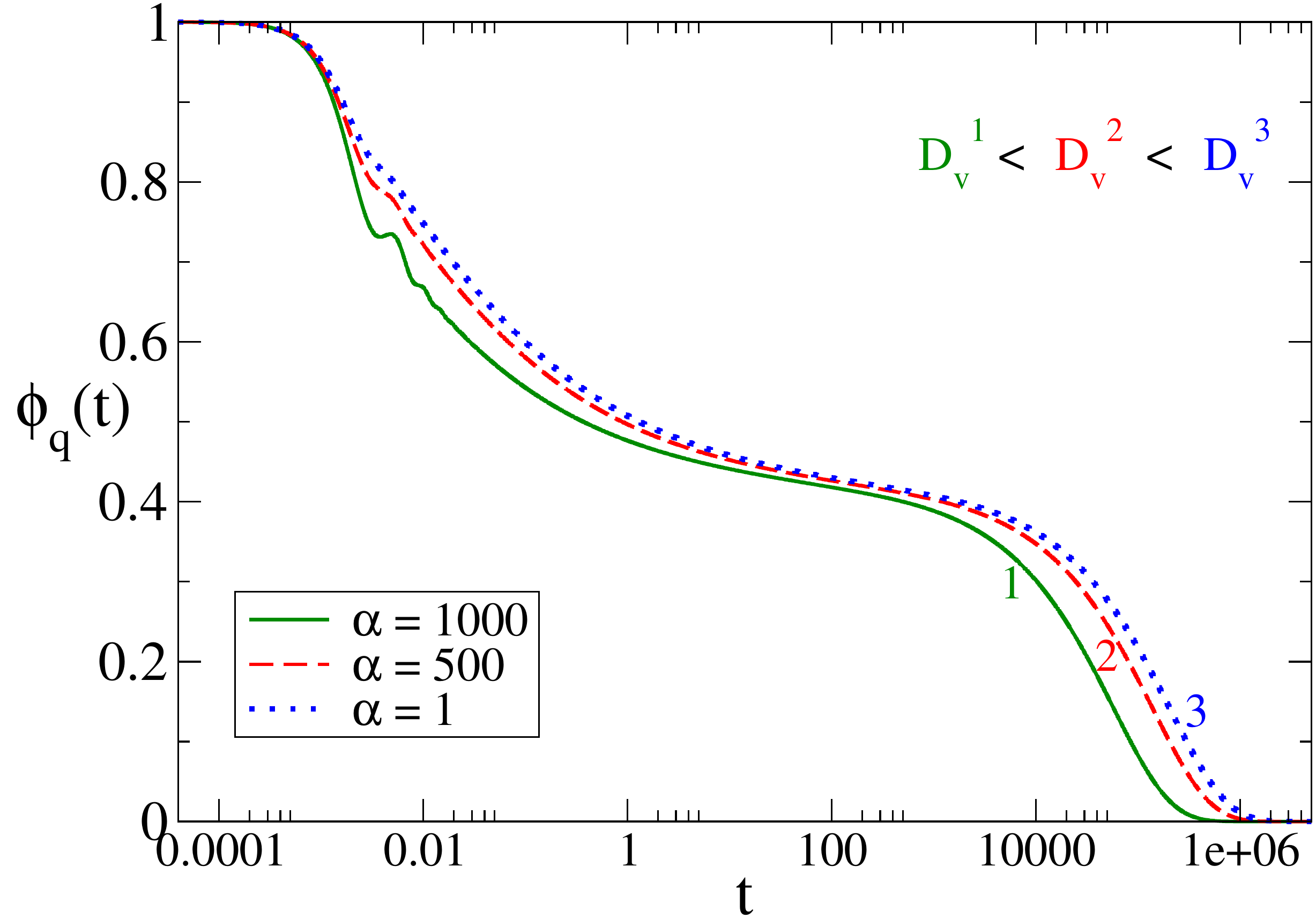}
\caption{\label{fig:phi_newton}Density correlation function $\phi_q(t)$ 
following Eq.~(\ref{Eq:eq_m_newton}) for $q = 4.2$ and packing fraction 
$\varphi=0.72449$ equivalent to $\varepsilon = (\varphi - 
\varphi_c)/\varphi_c \simeq 0.0002$, when $\langle \mathbf{v}^2 \rangle = 
2k_\text{B}T_\text{eff} = 1010$, $\alpha$ values presented in the legend 
and from Eq.~(\ref{Eq:v2}) $D^1_v = 88385.66$, $D_v^2 = 501096.48$ and 
$D_v^3 = 800608.13$. The higher the activity of the system (larger 
$\alpha$ and smaller $D_v$) the sooner the correlation function decays.}
\end{figure}

\begin{figure}[hbtp]
\centering
\includegraphics[width=0.9\columnwidth]{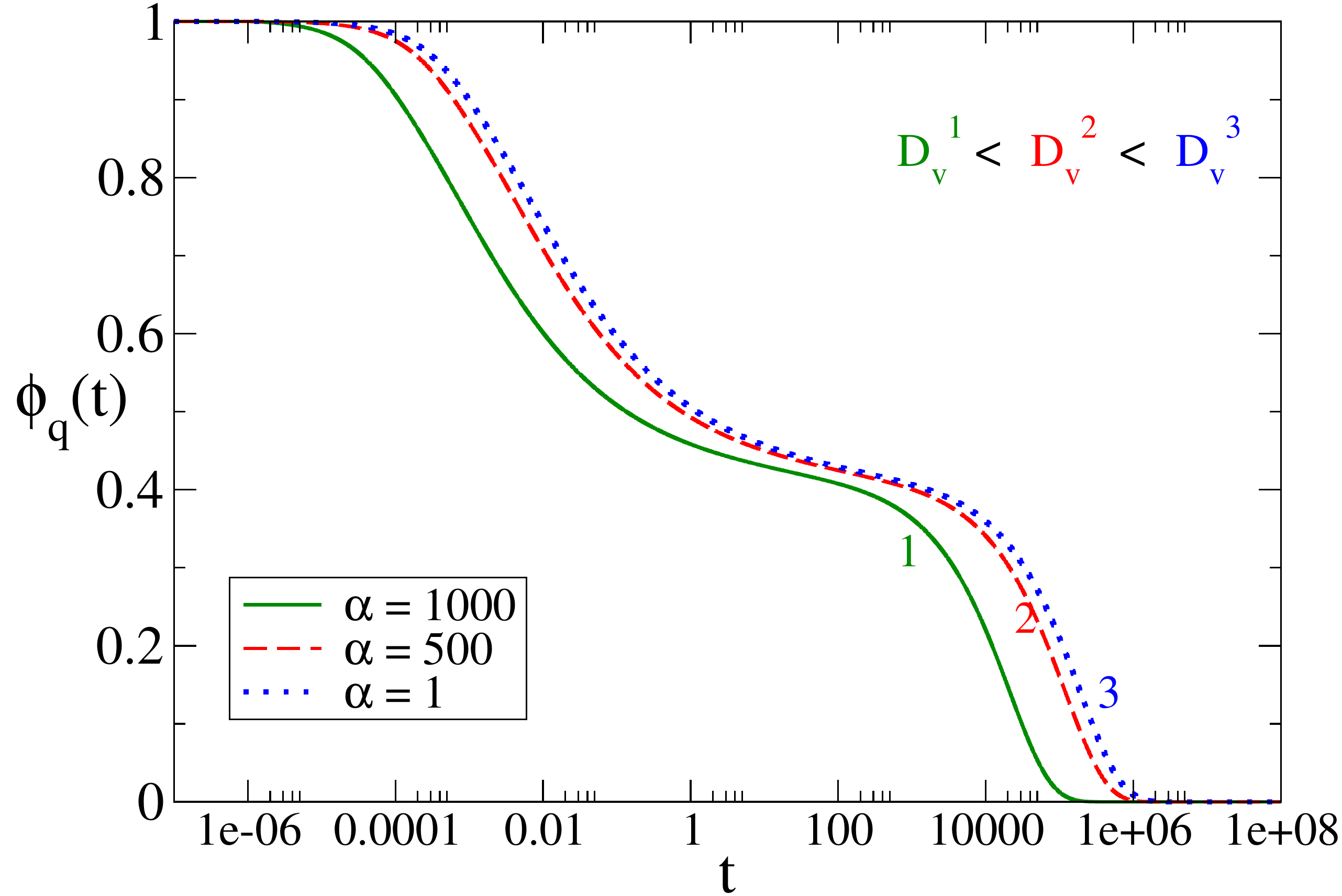}
\caption{\label{fig:phi_newton0}Density correlation function $\phi_q(t)$ 
following Eq.~(\ref{Eq:eq_m_newton0}) for overdamped motion for $q = 4.2$ 
and packing fraction $\varphi = 0.72449$ equivalent to $\varepsilon = 
(\varphi - \varphi_c)/\varphi_c \simeq 0.0002$, when $\langle \mathbf{v}^2 
\rangle = 2k_\text{B}T_\text{eff} = 1010$, $\alpha$ values presented in 
the legend and from Eq.~(\ref{Eq:v2}) $D^1_v = 88385.66$, $D_v^2 = 
501096.48$ and $D_v^3 = 800608.13$. The higher the activity of the system 
(larger $\alpha$ and smaller $D_v$) the sooner the correlation function 
decays.}
\end{figure}

\noindent%
As the kernel $m^\text{mct}_q$ obtained here is the same as in the case of 
normal Brownian motion, the glass transition packing fraction will also 
not change. But the damping coefficient in both Eq.~(\ref{Eq:eq_m_newton}) 
and (\ref{Eq:eq_m_newton0}) is different from the equilibrium case. The 
input to the equations of motions is the static structure factor $S_q$. In 
the next section, we shall use the ITT formalism to investigate the 
possible changes in the structure factor as a result of the nonequilibrium 
situation. For now, we use the Baus-Colot \cite{Baus1986,Bayer2007} 
analytical expression for the structure factor of the hard-sphere system 
in two dimensions (hard disks) to solve the equations of motion. The glass 
transition happens at the critical packing fraction $\varphi_c = 0.72464$. 
We have used $500$ grid points in the range $q_\text{min} = 0.04$ to 
$q_\text{max} = 39.96$ with $\Delta q=0.08$ to solve the integral equations.

We choose the temperature $\langle \mathbf{v}^2 \rangle = 2k_\text{B} 
T_\text{eff} = 1010$ and we consider three pairs of parameters $(\alpha, 
D_v) = (1000, 88385.66), (500, 501096.48), (1, 800608.13)$ with the 
mentioned temperature. We use Eq.~(\ref{eq:n_active}) to obtain the 
probability of finding active particle in the system for these three 
different pairs of parameters. The resulting values are $P_\text{active} = 
0.0006$, $0.2767$ and $0.4956$ for $(\alpha, D) = $ $(1, 800608.13)$ 
,$(500, 501096.48)$ and $(1000, 88385.66)$ respectively.  In 
Fig.~\ref{fig:phi_newton}, the solution of Eq.~(\ref{Eq:eq_m_newton}) for 
$\phi_q(t)$ with the packing fraction $\varphi = 0.72449$ in the liquid 
state and close to transition is presented for the three aforementioned 
pairs of $(\alpha,D_v)$. The higher the probability of finding active 
particles in the system, the smaller the time that the correlation 
function decays to zero. The same behavior is observed for the overdamped 
case. The solution of Eq.~(\ref{Eq:eq_m_newton0}), considering the same 
input, is shown in Fig.~\ref{fig:phi_newton0}.

\begin{figure}[htb] \centering 
\includegraphics[width=0.9\columnwidth]{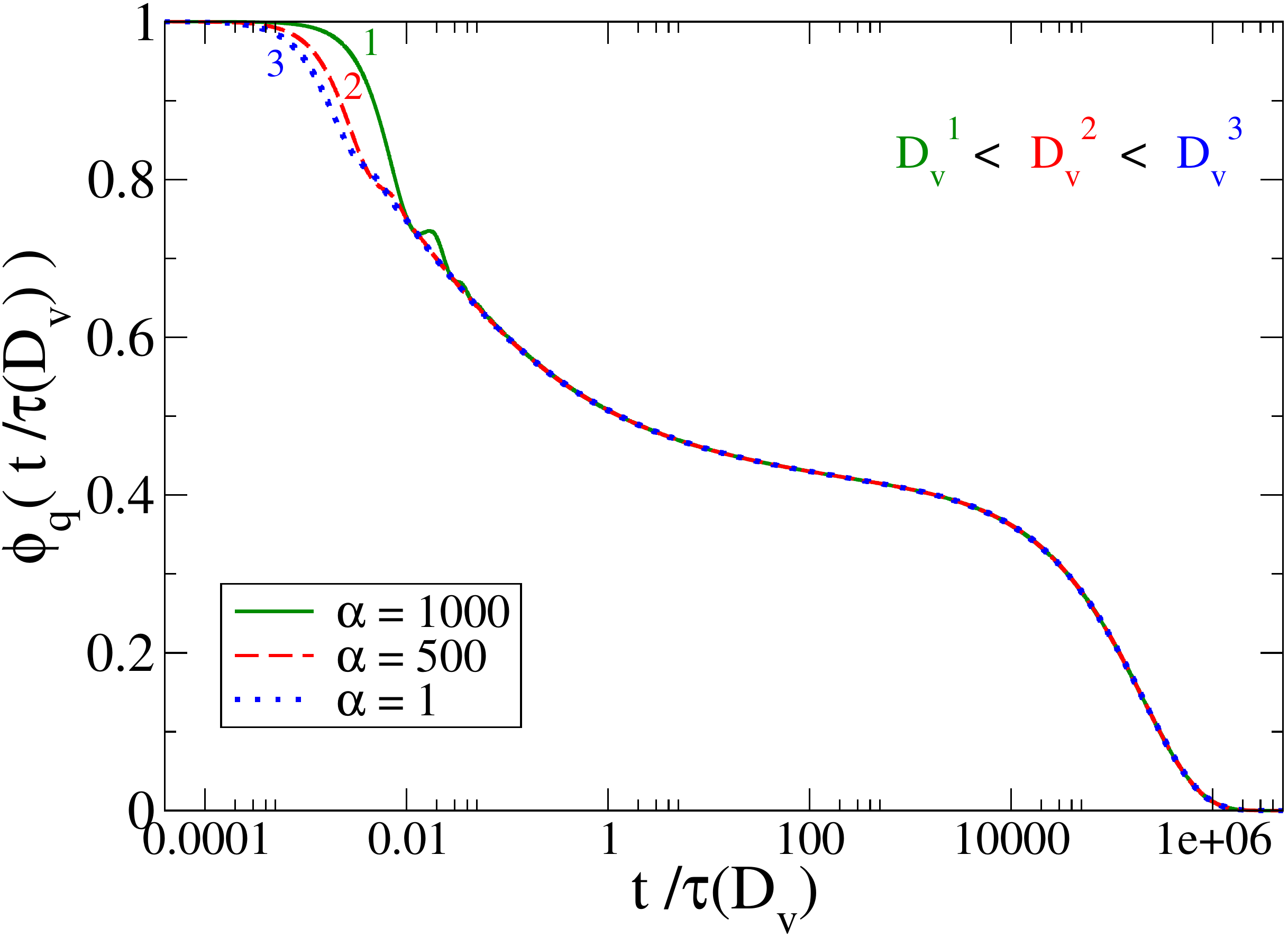} 
\caption{\label{fig:newton_scaled} Scaled density correlation function 
$\phi_q(\tilde{t})$ according to Eq.~(\ref{Eq:scale_newton}) for $q = 4.2$ 
and packing fraction $\varphi = 0.72449$ equivalent to $\varepsilon = 
(\varphi - \varphi_c)/\varphi_c \simeq 0.0002$, when $\langle \mathbf{v}^2 
\rangle = 2k_\text{B}T_\text{eff} = 1010$. $\tau(D_v)=1$ for $(\alpha, 
D^3_v) = (1, 800608.13)$ $\tau(D_v)=1$, $\tau(D_v^2) = 0.681$ for 
$(\alpha, D^2_v) = (500, 501096.48)$, and $\tau(D_v^1) = 0.283$ for 
$(\alpha, D^1_v) = (1000, 88385.66)$. }
\end{figure}

Since introducing the velocity-dependent friction does not cause any 
change in the memory kernel, the activity in the presented model does not 
effect directly the glass transition packing fraction which indicates that 
activity does not melt the glass. However it can shift the correlation 
function in the way that for a constant temperature and below the glass 
transition packing fraction, the higher the percentage of active 
particles in the system, the smaller is the time that the correlation 
function decays to zero. For a better comparison we use the second scaling 
law ($\alpha$-scaling) \cite{Goetze2009}. We scale the time in the 
correlation functions shown in Fig.~\ref{fig:phi_newton}, in a way that 
all three correlations fall on top each other in the long time regime. The 
scaling follows
\begin{equation}\label{Eq:scale_newton}
 \phi_q(\tilde{t})= \phi_q\left(\frac{t}{\tau(D_v)}\right),
\end{equation}
where $\tau(D_v)$ is the scaling time depending on $D_v$. For the 
correlation function corresponding to $(\alpha, D_v) = (1000, 88385.66)$, 
we find $\tau(D_v)=0.283$; for $(\alpha, D_v) = (500,501096.48)$, 
$\tau(D_v) = 0.681$; and for $(\alpha,D_v) =(1, 800608.13)$, the time 
scale is $\tau(D_v) = 1$. The scaled correlation functions are shown in 
Fig.~\ref{fig:newton_scaled}. Except for the short time dynamics, the 
correlation functions fall on top of each other. One should have in mind 
that the scaling time $\tau(D_v)$ will not diverge as function of $D_v$, 
since the glass transition packing fraction is not dependent on activity 
and for packing fractions below $\varphi_c$, the correlation function will 
always decay to zero. Since the structure factor is the static input to 
the equations, small changes in structure factors can change the 
mode-coupling predictions about the glass transition drastically. In the 
next section, we shall study the possible changes in the structure factor. 

\section{Integration Through Transients}\label{itt}

If the distribution function $f$ in Eq.~(\ref{Eq:dis_total_R}), was a 
stationary solution of the Fokker-Planck equation (\ref{Eq:fk}), 
substituting $f$ inside the Fokker-Planck equation would result in 
$\partial f/\partial t = 0$. But as mentioned before, $f$ is not a general 
solution of the Fokker-Planck equation and is only an estimate of the 
stationary distribution. Replacing $f$ in the Fokker-Planck equation 
yields $\partial f/\partial t=\Lambda f$ where $\Lambda$ follows 
Eq.~(\ref{Eq:Lambda}). From Eq.~(\ref{Eq:rond_f}) it is seen that $f$ will 
be a solution of the Fokker-Planck equation under the condition that 
$\mathbf{F}_i=0$. In the situation $\mathbf{F}_i\neq0$ with normal 
friction, the equilibrium structure factor $S_q$ is justified.
We use this fact here and assume when $t<0$ the 
interaction forces $\mathbf{F}_i$ are switched off, and at $t=0$ we switch 
on the interaction forces. Therefore we can refer to $f$ as the stationary 
distribution function when $t<0$. 
Using the ITT formalism we are able to 
evaluate the time dependence of the distribution function as
\begin{equation}
 f(\mathbf{\Gamma},t) =
\begin{cases}
f(\mathbf{\Gamma}),&  t\leq 0\\
e^{\Lambda t} f(\mathbf{\Gamma}),&  t>0.
 \end{cases}
\end{equation}
Here $f$ follows Eq.~(\ref{Eq:dis_total_R}) and $f(\mathbf{\Gamma},t)$ is  
the time dependent distribution function. One can write \cite{Fuchs2009}
\begin{equation}
 e^{\Lambda t}=1+\int_0^t dt'~e^{\Lambda t'} \Lambda,
\end{equation}
therefore when $t \rightarrow \infty$ according to the Integration Through 
Transients (ITT) formalism \cite{Fuchs2009}
\begin{equation}
\begin{split}
 \int &\text{d}\mathbf{\Gamma} f(\mathbf{\Gamma},t) 
\rho_\mathbf{q}^\ast~\rho_\mathbf{q}\\
 &= \int \text{d}\mathbf{\Gamma} f(\mathbf{\Gamma}) 
\rho_\mathbf{q}^\ast~\rho_\mathbf{q}+\int d\mathbf{\Gamma} 
 \int_0^\infty \text{d}t~\rho_\mathbf{q}^\ast~\rho_\mathbf{q}
 e^{\Lambda t} \Lambda f(\mathbf{\Gamma})
 \end{split}
 \end{equation}
or
 \begin{equation}\label{eq:itt_in_mymodel}
 \begin{split}
 NS^s_q&=NS_q+ \int_0^\infty \text{d}t \int d\mathbf{\Gamma}~\Lambda
 f(\mathbf{\Gamma}) e^{-\Lambda t}\rho_\mathbf{q}^\ast~\rho_\mathbf{q}.
\end{split}
\end{equation}
Here, $S^s_q$ is the structure factor in the stationary state which is 
reached for $t\rightarrow \infty$. We assume that we can replace 
$-\Lambda$ with $i\mathcal{L}$
\begin{equation}\label{eq:lambda_L_rel}
 e^{-\Lambda t} \rho_\mathbf{q}^\ast~\rho_\mathbf{q}=e^{i\mathcal{L} 
t}\rho_\mathbf{q}^\ast~\rho_\mathbf{q}.
\end{equation}
Using the projection operator $\mathcal{Q} = 1 - 
\sum_\mathbf{q}\rho_\mathbf{q} \langle\rho^\ast_\mathbf{q}|\dots\rangle/N 
S_q$, from Eq.~(\ref{eq:itt_in_mymodel}) and (\ref{eq:lambda_L_rel}) we 
arrive at
\begin{equation}
N S^{s}_q = N S_q+\int_0^\infty \text{d}t \langle \Lambda \mathcal{Q} 
e^{i \mathcal{Q}\mathcal{L} \mathcal{Q} t}\mathcal{Q} 
\rho_\mathbf{q}^\ast~\rho_\mathbf{q}\rangle.
\end{equation}
Using the mode coupling approximation
\begin{equation}\label{Eq:ITT_k}
\begin{split}
\langle \Lambda& \mathcal{Q P} e^{i\mathcal{L} t}\mathcal{P Q} 
\rho_\mathbf{q}^\ast~\rho_\mathbf{q}\rangle\\
&=\sum_{\textbf{k}< \textbf{p}} \frac{\langle \Lambda\mathcal{Q}|
\rho_{\textbf{k}} \rho_{\textbf{p}}\rangle 
\langle \rho^{\ast}_{\textbf{k}}\rho^{\ast}_{\textbf{p}}| 
\exp{(i\mathcal{L}t)}\rho_{\textbf{k}} \rho_{\textbf{p}}\rangle
\langle\rho^{\ast}_{\textbf{k}} \rho^{\ast}_{\textbf{p}}| \mathcal{Q} 
\rho_\mathbf{q}^\ast \rho_\mathbf{q}\rangle}{\langle 
\rho^{\ast}_{\textbf{k}} \rho^{\ast}_{\textbf{p}}|\rho_{\textbf{k}} 
\rho_{\textbf{p}}\rangle^2}.
\end{split}
\end{equation}
From Eq.~(\ref{Eq:Lambda})
\begin{equation}
\begin{split}
\langle \Lambda \rangle= \int \text{d}\mathbf{\Gamma} f(\mathbf{\Gamma}) 
\Lambda &= \left( 3\langle \mathbf{v}^2 \rangle-\frac{\xi^2}{\langle 
\mathbf{v}^2 \rangle}-\alpha\right)\\
&= \left( 3\langle \mathbf{v}^2 \rangle-\frac{2 D_v}{\langle \mathbf{v}^2 
\rangle}-\alpha\right),
\end{split}
\end{equation}
where $f(\mathbf{\Gamma})$ follows Eq.~(\ref{Eq:dis_total_R}). Also,
\begin{equation}\label{Eq:ITT_k_1}
 \langle \Lambda \mathcal{Q}|\rho_{\textbf{k}} \rho_{\textbf{p}}\rangle 
= N\delta_{-\mathbf{k},\mathbf{p}}\langle \Lambda \rangle S_k,
\end{equation}
and
\begin{equation}\label{Eq:ITT_k_2}
\begin{split}
 \langle&\rho^{\ast}_{\textbf{k}} \rho^{\ast}_{\textbf{p}}| \mathcal{Q} 
 \rho^\ast_\mathbf{q}\rho_\mathbf{q}\rangle\\
 &= \langle\rho^{\ast}_{\textbf{k}}
 \rho^{\ast}_{\textbf{p}}| \rho_\mathbf{q}^\ast\rho_\mathbf{q}\rangle - 
\sum_\mathbf{q} \frac{\langle\rho^{\ast}_{\textbf{k}} 
\rho^{\ast}_{\textbf{p}}|\rho_\mathbf{q}\rangle \langle 
\rho_\mathbf{q}^\ast|\rho_\mathbf{q}^\ast\rho_\mathbf{q} \rangle}{\langle 
\rho_\mathbf{q}| \rho^\ast_\mathbf{q}\rangle}\\
&=\delta_{-\mathbf{k},\mathbf{p}}\delta_{\mathbf{q},\mathbf{k} 
+ \mathbf{p}} N^2 S_k S_q - \sum_\mathbf{q}\frac{N^2 
\delta_{\mathbf{q},\mathbf{k} + \mathbf{p}} \delta_{\mathbf{q},\mathbf{q}+
\mathbf{q}} S_k S_p S_q S_q^3}{N S_q}\\
&=\delta_{-\mathbf{k},\mathbf{p}} 
\delta_{\mathbf{q},\mathbf{k}+\mathbf{p}}N^2S_k(1-S_k).
\end{split}
\end{equation}
Substitution of Eq.~(\ref{Eq:ITT_k_1}) and (\ref{Eq:ITT_k_2}) into 
(\ref{Eq:ITT_k}) results in
\begin{equation}
 \langle \Lambda \mathcal{Q P} e^{i\mathcal{L} t}\mathcal{P Q} 
\rho_\mathbf{q}^\ast~\rho_\mathbf{q}\rangle = \frac{1}{2}
\langle \Lambda \rangle N (1-S_k) \phi^2_k(t).
\end{equation}
Therefore,
\begin{equation}
S^{s}_q =S_q+ \frac{1}{2} \langle \Lambda \rangle
 (1-S_q)\int_0^\infty \phi^2_q(t)\text{d}t, 
\end{equation}
or finally,
\begin{equation}\label{Eq:itt_sqs}
S^{s}_q = S_q + \frac{1}{2} \left( 3\langle \mathbf{v}^2 \rangle 
 	- \frac{2 D_v}{\langle \mathbf{v}^2 \rangle}-\alpha\right)
 (1-S_q)\int_0^\infty \phi^2_q(t)\text{d}t. 
\end{equation}

This equation is very similar to what Farage \textit{et al.} 
\cite{Farage2014} have obtained.

We obtain the correlation function 
$\phi_q(t)$ from Eq.~(\ref{Eq:eq_m_newton0}) and substitute it into 
Eq.~(\ref{Eq:itt_sqs}) to calculate $S^{s}_q$. The integral 
$\int_0^\infty \phi^2_q(t)\text{d}t$ becomes 
infinitely large at the glass transition, therefore we are able to 
calculate $S^{s}_q$ only when we are sufficiently away from the glass 
transition and inside the liquid state. The other necessity for 
Eq.~(\ref{Eq:itt_sqs}) to result in a reasonable $S^{s}_q$ is that the 
effective temperature should be sufficiently low. In other words, solving 
Eq.~(\ref{Eq:itt_sqs}) requires that the perturbations are adequately 
small.

\begin{figure}[htb] \centering
\includegraphics[width=0.9\columnwidth]{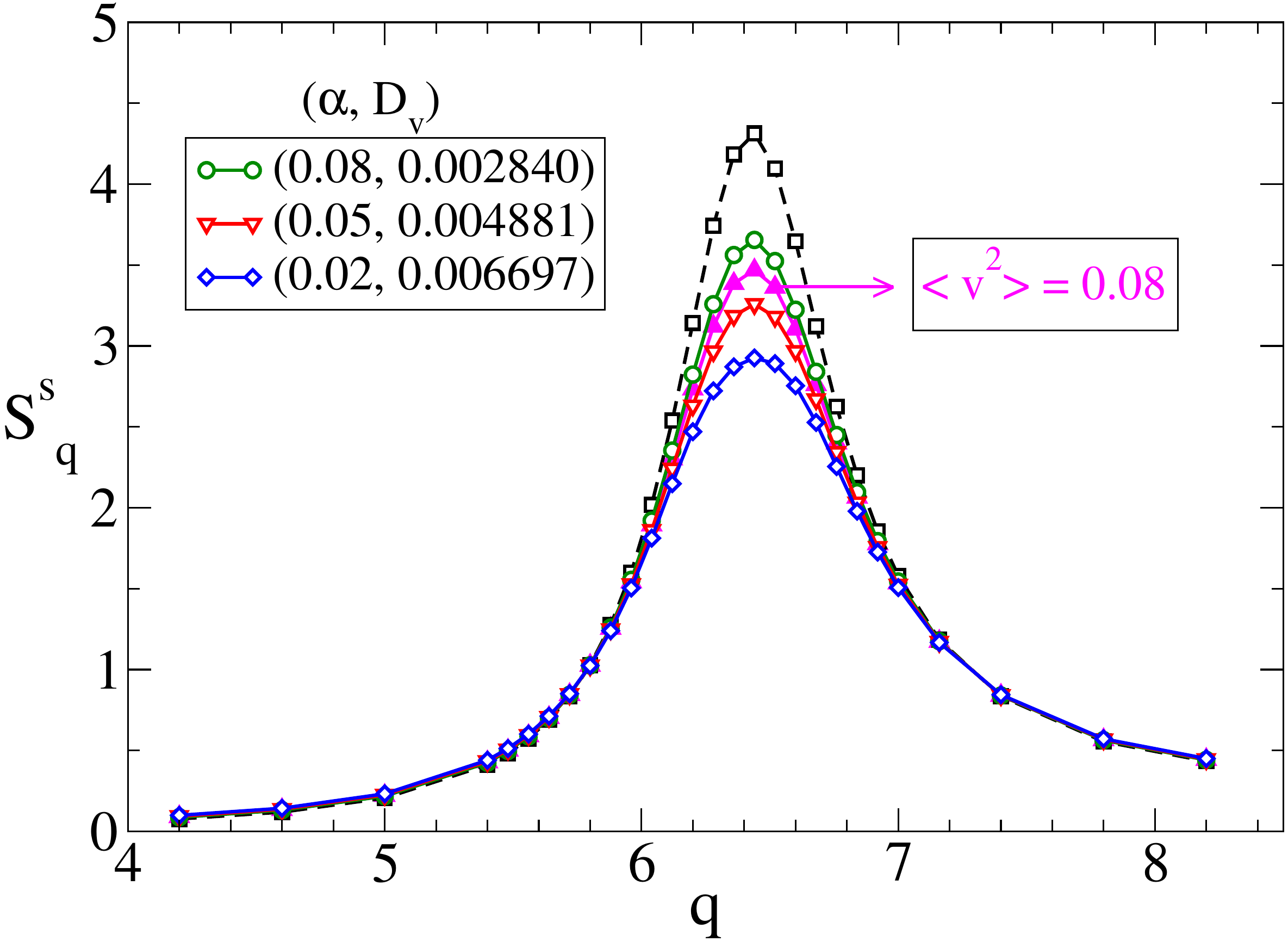}
\caption{\label{fig:Sqs_itt}Structure factor $S^s_q$ for values around the 
first peak, calculated via Eq.~(\ref{Eq:itt_sqs}), for three pairs of 
$(\alpha, D_v)$ as indicated in the legends when $\langle 
\mathbf{v}^2\rangle = 2k_\text{B}T_\text{eff} = 0.1$ and also for
$(\alpha', D'_v) = (0.04, 0.003124)$ when $\langle 
\mathbf{v}^2\rangle = 2k_\text{B}T_\text{eff} = 0.08$ shown with 
filled upwards triangles. The Baus-Colot equilibrium 
structure factor $S_q$ is shown with squares. $\varepsilon = 
(\varphi_c-\varphi)/\varphi_c \simeq 0.0215$.}
\end{figure}

For $\varepsilon = (\varphi_c-\varphi)/\varphi_c\simeq 0.0215$ and 
$\langle \mathbf{v}^2\rangle = 2k_\text{B}T_\text{eff} = 0.1$ we have 
solved Eq.~(\ref{Eq:eq_m_newton0}) for three pairs of $(\alpha, D_v) = 
(0.08, 0.00284), (0.05, 0.004881)$, and $(0.02, 0.006697)$. As we 
discussed in section \ref{percent}, the higher the $\alpha$ (the 
smaller the $D_v$), the higher is the percentage of active particles 
in the system. Therefore these three pairs correspond to monotonically 
decreasing fractions of active particles, with all three pair at the 
same effective temperature. For having a good comparison we also 
introduce a fourth pair $(\alpha', D'_v)$ at a smaller effective temperature
than the aforementioned three pairs, but the same fraction of active 
particles as in $(0.05, 0.004881)$. We chose the effective temperature 
for the forth term to be $2k_\text{B}T_\text{eff} = 0.08$. According to 
Eq.~(\ref{eq:n_active}) for the $(\alpha', D'_v)$ to have the same 
$P_\text{active}$ as $(0.05, 0.004881)$ has, $\alpha'/\sqrt{D'_v}$ must 
be equal to $0.05/\sqrt{0.004881}$. This together with the condition that
$2k_\text{B}T_\text{eff} = 0.08$ results into $(\alpha', D'_v) = (0.04, 0.003124)$.
For solving Eq.~(\ref{Eq:eq_m_newton0}) we use the Baus-Colot analytical 
expression for the structure factor $S_q$ of the hard-sphere system in two 
dimensions \cite{Baus1986,Bayer2007}. For every $q$ value, replacing $\phi_q(t)$ 
in Eq.~(\ref{Eq:itt_sqs}) and calculating the integral $\int_0^\infty 
\phi^2_q(t)\text{d}t$ results in the $S^{s}_q$. We show the $S^{s}_q$ 
values around the first peak, in Fig.~\ref{fig:Sqs_itt}. For the three 
pairs with the same effective temperature, one can observe that with 
decreasing $\alpha$, the peak value of the $S^s_q$ decreases too. This 
is different from \cite{Farage2014}. Here, we model the activity with 
velocity-dependent friction which is isotropic and does not have any 
rotational or directional dependence. But we are adding an additional 
constraint to the system. This additional constraint is $D_v$ related 
to the percentage of active particles in the system. The higher is that 
percentage (the smaller is the $D_v$), the more ordered the system becomes 
and the higher is the peak value of the structure factor. A comparison between 
the structure factor peak of $(0.05, 0.004881)$ and $(\alpha', D'_v) = (0.04, 0.003124)$ 
shows that as we may expect, although these two curves correspond to the 
same percentage of activity in the system, since the temperature is lower 
when $(\alpha', D'_v) = (0.04, 0.003124)$ the structure factor peak has 
larger peak value.

In general, the structure factors $S^s_q$ are less pronounced than the 
equilibrium Baus-Colot structure factor. In other systems, e.g. colloidal 
suspensions with short-ranged attractive interactions \cite{Dawson2000}, it 
has been shown that a decrease in the structure factor peak value yields 
an increase of the packing fraction for the glass transition according to 
MCT equations. Therefore we conclude that the less pronounced peak in the 
structure factors $S^s_q$ would result in higher transition packing 
fractions.
The change in the structure factor first peak due to activity has been 
reported before. Ni et al. \cite{Ni2013} have shown by simulation that the 
structure factor peak value of an active system of self-propelled hard spheres 
will reduce by increasing activity and the glass transition shifts to higher 
packing fractions. The same result for the structure factor was obtained earlier in 
a simulated system of motorized particles \cite{Loi2008}. Szamel et al. \cite{Szamel2015pre} 
also show the changes in structure factor and transition point in response to 
increasing activity although those changes are not monotonic.

\section{Conclusion}

We analyzed the glassy dynamics of a system in which slow particles are 
accelerated and fast particles are damped, by means of extending 
mode-coupling theory to nonequilibrium situations. We have approximated 
the distribution function by the solution of the Fokker-Planck equation 
for a noninteracting system. In that case, the activity does not affect 
the glass transition directly in the memory kernel as in the case for 
granular matter \cite{Kranz2010,Kranz2013}. However, in the present system 
activity leads to a modification of the static structure factor as shown 
above by employing the ITT formalism together with a factorization 
approximation, cf. Fig.~\ref{fig:Sqs_itt}. In general the structure factor 
peak values for the considered active systems are smaller than the 
equilibrium Baus-Colot structure factor peak value. Hence, one expects a 
shift of the glass transition packing fractions in the active systems 
towards higher values in comparison to the equilibrium case. Such a trend 
was observed in the numerical simulation results \cite{Ni2013} for a 
related active system for both the glass transition density as well as the 
variation of the static structure factor with activity, lending support to 
the a priory uncontrolled approximations used in the MCT and ITT 
calculations.

\section{Acknowledgment}
We thank W.T. Kranz for reading the manuscript critically. We acknowledge 
financial support from DAAD and DFG under FG1394.

\appendix
\section{Noise Terms}\label{sec:appA}

As mentioned in section \ref{liou}, both time evolution operators 
$i\mathcal{L}$ and $i \mathcal{L}^{\dagger}$ contain the term 
$\boldsymbol{\xi} R_i(t)\cdotp\frac{\partial}{\partial\mathbf{p}_i}$. 
Since $\boldsymbol{\xi}R_i(t)$ is a stochastic force, the time evolution, 
would be different for every realization. Therefore we take an average 
over the noise. Here we review the calculation of these averages in detail 
following \cite{Kubo1991}.  We assume $\frac{\text{d} 
B(\Gamma(t))}{\text{d} t} = i\mathcal{L}_1~B(\Gamma(t)) = 
-\boldsymbol{\xi} R_i(t)\cdotp\frac{\partial}{\partial\mathbf{p}_i} 
B(\Gamma(t))$ therefore
\begin{equation}\label{Eq:B}
 B(t+\Delta t)-B(t) = \int_t^{t + \Delta t} 
	i\mathcal{L}_1 B(t_1) \text{d} t_1.
\end{equation}
We substitute $B$ from Eq.~(\ref{Eq:B}) into itself and drop $B$ from both 
sides of the equation, $-\boldsymbol{\xi} R_i(t)\cdotp 
\frac{\partial}{\partial\mathbf{p}_i}$ is equal to
\begin{equation}
\begin{split}
& \lim_{\Delta t \to 0} \frac{1}{\Delta t} \left[ \int_{t}^{t+\Delta t} 
- \boldsymbol{\xi} R_i(t_1)\cdotp\frac{\partial}{\partial\textbf{p}_i} 
\text{d}t_1 \right. \\
& \left. + \int_{t}^{t+\Delta t}\int_{t}^{t_1} \left(\boldsymbol{\xi} 
R_i(t_1)\cdotp\frac{\partial}{\partial\textbf{p}_i}\right)
 \left(\boldsymbol{\xi} R_i(t_2)\cdotp
\frac{\partial}{\partial\textbf{p}_i}\right) \text{d}t_1 \text{d}t_2 
\right].
\end{split}
\end{equation}
Since the time scale of $R_i(t)$ is much shorter than the phase variables, 
we can choose $\Delta t$ long enough that we can replace the terms inside 
the integrals by their averages
\begin{equation}
\begin{split}\label{Eq:random_force_ave}
&-\boldsymbol{\xi} R_i(t)\cdotp\frac{\partial}{\partial\mathbf{p}_i}\\
&=\lim_{\Delta t \to 0} \frac{1}{\Delta t}  
	\left[ \int_{t}^{t+\Delta t} - \langle \boldsymbol{\xi}
 R_i(t_1)\cdotp\frac{\partial}{\partial\textbf{p}_i}\rangle 
	\text{d}t_1 \right. \\
&\left.  + \int_{t}^{t+\Delta t}\int_{t}^{t_1} 
\left\langle\left(\boldsymbol{\xi} R_i(t_1)\cdotp 
\frac{\partial}{\partial\textbf{p}_i}\right) 
\left(\boldsymbol{\xi} R_i(t_2)\cdotp 
\frac{\partial}{\partial\textbf{p}_i}\right)\right\rangle \text{d}t_1 
\text{d}t_2 \right].
\end{split}
\end{equation}
According to Eq.~(\ref{Eq:ave_R}) the first part of the right hand side of 
Eq.~(\ref{Eq:random_force_ave}) is zero and
\begin{equation}
\begin{split}\label{Eq:rfa2}
-\boldsymbol{\xi}& R_i(t)\cdotp\frac{\partial}{\partial\mathbf{p}_i}\\
& = \lim_{\Delta t \to 0} \frac{\xi^2}{\Delta t}   \int_{t}^{t+\Delta t}
\int_{t}^{t_1} \langle R_i(t_1) R_i(t_2)\rangle 
\frac{\partial^2}{{\partial p_i}^2} \text{d}t_1 \text{d}t_2 \\
&= \lim_{\Delta t \to 0} \frac{\xi^2}{2 \Delta t}  \int_{t}^{t+\Delta t} 
\frac{\partial^2}{{\partial p_i}^2} \text{d}t_1  \\
&= \frac{1}{2} \xi^2 \frac{\partial^2}{{\partial p_i}^2},
\end{split}
\end{equation}
where we have used the property of the Dirac delta 
$\int_t^{t_1}~\delta(t_1-t_2)~\text{d}t_2 = 1/2$ where $t<t_2<t_1$.

\section{Velocity Integrals}\label{sec:appB}

\noindent%
Here we calculate the integrals in Eq.~(\ref{Eq:1_C}), (\ref{Eq:v2}), 
(\ref{Eq:v4}) and (\ref{Eq:v6}) as
\begin{equation}
\begin{split}
 \frac{1}{C}& = 2\pi\int_0^\infty e^{-\left(\frac{v^4}{4D_v}-\frac{\alpha 
v^2}{2D_v}\right)} v\text{d}v\\
 &= 2\pi 
\sqrt{D_v}~e^{\frac{\alpha^2}{4D_v}} 
\int_{\frac{-\alpha}{2\sqrt{D_v}}}^\infty e^{-U^2} \text{d}U,
 \end{split}
\end{equation}
where $U = \frac{v^2}{2\sqrt{D_v}}-\frac{\alpha }{2\sqrt{D_v}}$. Therefore 
\begin{equation}\label{Eq:1_C_A}
\begin{split}
 \frac{1}{C} &= 2\pi \sqrt{D_v}~e^{\frac{\alpha^2}{4D_v}}\left(\int_{ 
\frac{-\alpha}{2\sqrt{D_v}}}^0 e^{-U^2}\text{d}U+\int_0^\infty 
e^{-U^2}\text{d}U\right)\\
 &= \pi \sqrt{\pi D_v}~ \exp\left(\frac{\alpha ^2}{4D_v}\right) 
\left[1+\text{erf}\left(\frac{\alpha}{2\sqrt{D_v}}\right) \right],
 \end{split}
\end{equation}
where we used the definition of the error function $\text{erf}(x) = 
\int_0^x e^{-t^2} \text{d}t$ and the integral $\int_0^\infty e^{-t^2} 
\text{d}t = \sqrt{\pi}/2$. Also
\begin{equation}
 \begin{split}
  \langle \mathbf{v}^2 \rangle =& 2\pi C e^{\frac{\alpha^2}{4D_v}} 
\int_0^\infty e^{-\left(\frac{v^2}{2\sqrt{D_v}} - \frac{\alpha 
}{2\sqrt{D_v}}\right)^2} v^2~v~\text{d}v\\
  =& 2\pi \sqrt{D_v}~e^{\frac{\alpha^2}{4D_v}}
\int_{\frac{-\alpha}{2\sqrt{D_v}}}^\infty e^{-U^2} 
2\sqrt{D_v}~(U+\frac{\alpha}{2\sqrt{D_v}})~\text{d}U\\
  =& 4\pi D_v~e^{\frac{\alpha^2}{4D_v}}\left(\int_{
\frac{-\alpha}{2\sqrt{D_v}}}^\infty 
\frac{\alpha}{2\sqrt{D_v}}~e^{-U^2}~\text{d}U\right.\\
  & \left.+\int_{\frac{-\alpha}{2\sqrt{D_v}}}^\infty U~e^{-U^2} 
\text{d}U\right).
 \end{split}
\end{equation}
The first integral is proportional to $1/C$ and the second integral can be 
calculated easily
\begin{equation}
 \int_{\frac{-\alpha}{2\sqrt{D_v}}}^\infty U~e^{-U^2} \text{d}U = 
\frac{1}{2}~e^{\frac{-\alpha^2}{4D_v}}.
\end{equation}
Therefore
\begin{equation}
 \langle \mathbf{v}^2 \rangle= \alpha+ 2 \sqrt{\frac{D_v}{\pi}}
 \exp\left(-\frac{\alpha ^2}{4D_v}\right){\left[1 + \text{erf}\left(
\frac{\alpha}{2\sqrt{D_v}}\right) \right]}^{-1}.
\end{equation}
This is different from the expression in \cite{Erdmann2000} by a minus 
sign in the exponent of $\exp\left(-\frac{\alpha^2}{4D_v}\right)$. We go 
ahead and use the same method as \cite{Erdmann2000,Stratonovich1967} to 
obtain $\langle \mathbf{v}^4 \rangle$ and also $\langle \mathbf{v}^6 
\rangle$,
\begin{equation}
\langle \mathbf{v}^4 \rangle = \frac{4D^2_v}{C^{-1}}
\frac{\partial^2}{\partial \alpha^2} (C^{-1}),
\end{equation}
where $C^{-1}$ follows Eq.~(\ref{Eq:1_C_A}). And
\begin{equation}
\langle \mathbf{v}^6 \rangle = \frac{8D^3_v}{C^{-1}}
\frac{\partial^3}{\partial \alpha^3} (C^{-1}).
\end{equation}
So
\begin{equation}\label{Eq:v4_A}
\langle \mathbf{v}^4\rangle=2D_v+  \alpha \langle \mathbf{v}^2\rangle, 
\end{equation}
and
\begin{equation}\label{Eq:v6_A}
\langle \mathbf{v}^6\rangle = 2\alpha D_v + (\alpha^2+4D_v) \langle 
\mathbf{v}^2\rangle. 
\end{equation}

\clearpage
\bibliographystyle{apsrev}
\bibliography{Bibliography} 

\end{document}